\documentclass[trans]{IEEEtran}
\usepackage{hyperref}
\usepackage{lettrine}
\usepackage{epsfig}
\usepackage{footnote}
\usepackage{cite}
\usepackage{float}
\usepackage{amssymb}
\usepackage{amsthm}
\usepackage{amsmath,mathabx}
\usepackage{color}
\usepackage{amsfonts}
\usepackage{epstopdf}
\usepackage{tabularx,tabulary}
\usepackage[font=footnotesize]{caption}
\usepackage{graphicx}
\usepackage{caption}
\usepackage{subcaption}
\usepackage{wrapfig}
\usepackage{dblfloatfix}    
\usepackage{bbm}
\usepackage{algorithm}
\usepackage[noend]{algpseudocode}
\makeatletter
\def\BState{\State\hskip-\ALG@thistlm}
\makeatother
\usepackage{bbold}

\newcommand{\vect}[1]{\boldsymbol{#1}}
\DeclareMathOperator{\sgn}{sgn}

\DeclareMathOperator{\argmin}{argmin}
\DeclareMathOperator{\argmax}{argmax}

\DeclareMathOperator{\erfc}{erfc}

\begin{document}
\title{End-to-End Performance Analysis of Underwater Optical Wireless Relaying and Routing Techniques Under Location Uncertainty}

\author{
Abdulkadir~Celik,~\IEEEmembership{Member,~IEEE,} Nasir~Saeed,~\IEEEmembership{Member,~IEEE,} Basem Shihada,~\IEEEmembership{Senior Member,~IEEE,} \\Tareq Y. Al-Naffouri,~\IEEEmembership{Senior Member,~IEEE,} and Mohamed-Slim~Alouini,~\IEEEmembership{Fellow,~IEEE}.

\thanks{Authors are with Computer, Electrical, and Mathematical Sciences and Engineering Division at King Abdullah University of Science and Technology (KAUST), Thuwal,  23955-6900, KSA. A part of this work was presented in IEEE WCNC 2018 in Barcelona, Spain \cite{Celik2018Modeling}.}
}
\maketitle
\thispagestyle{empty}
\pagestyle{empty}

\begin{abstract}
On the contrary of low speed and high delay acoustic systems, underwater optical wireless communication (UOWC) can deliver a high speed and low latency service at the expense of short communication ranges. Therefore, multihop communication is of utmost importance to improve degree of connectivity and overall performance of underwater optical wireless networks (UOWNs). In this regard, this paper investigates relaying and routing techniques and provides their end-to-end (E2E) performance analysis under the location uncertainty. To achieve robust and reliable links, we first consider adaptive beamwidths and derive the divergence angles under the absence and presence of a pointing-acquisitioning-and-tracking (PAT) mechanism. Thereafter, important E2E performance metrics (e.g., data rate, bit error rate, transmission  power, amplifier gain, etc.) are obtained for two potential relaying techniques; decode \& forward (DF) and optical amplify \& forward (AF). We develop centralized routing schemes for both relaying techniques to optimize E2E rate, bit error rate, and power consumption. Alternatively, a distributed routing protocol, namely \underline{Li}ght \underline{Pa}th \underline{R}outing (LiPaR), is proposed by leveraging the range-beamwidth tradeoff of UOWCs. LiPaR is especially shown to be favorable when there is no PAT mechanism and available network information. In order to show the benefits of multihop communications, extensive simulations are conducted to compare different routing and relaying schemes under different network parameters and underwater environments.  

 \end{abstract}
\IEEEpeerreviewmaketitle

\section{Introduction}
\label{sec:intro}

\lettrine{T}{he} recent demand on high quality of service communications for commercial, scientific and military applications of underwater exploration necessitates a high data rate, low latency, and long-range underwater networking solutions \cite{Celik2018Modeling}. Fulfilling these demands is a formidable challenge for most of the electromagnetic frequencies due to the highly attenuating aquatic medium. Therefore, acoustic systems have received considerable attention in the past decades. As a result of the surface-induced pulse spread and frequency-dependent attenuation, acoustic communication data rates are restrained to around tens of kbps for ranges of a kilometer, and less than kbps for longer ranges \cite{saeed2018survey}. Moreover, the low propagation speed of acoustic waves (1500 m/s) induces a high latency \cite{AKYILDIZ2005257}, especially for long-range applications where real-time communication and synchronization are challenging. 

Alternatively, underwater optical wireless communication (UOWC) has the virtues of supporting high speed connections in the order of Gbps \cite{Oubei:15,Shen:16}, providing low latency as a result of the high propagation speed of light in the aquatic medium ($\approx 2.55 \times 10^8$ m/s) \cite{Ballal2015low}, and enhancing security thanks to the point-to-point links \cite{Kaushal2016underwater}. Nevertheless, the reachable UOWC range is delimited by severe underwater channel impairments of absorption and scattering. The absorption effect refers to the energy dissipation of photons while being converted into other forms (e.g., heat, chemical, etc.) along the propagation path. The scattering is regarded as the deflection of the photons from its default propagation path, which is caused either by water particles of size comparable to the carrier wavelength (i.e., diffraction) or by constituents with different refraction index (i.e., refraction). Therefore, the relation between absorption and scattering primarily characterizes the fundamental tradeoff between range and beam divergence angle. That is, a collimated light beam can reach remote receivers within a tight span whereas a wider light beam can communicate with nearby nodes in a broader span. Accordingly, the range-beamdwidth tradeoff primarily determines the connectivity and outage in underwater optical wireless networks (UOWNs) \cite{Saeed2018Connectivity}. 

The directed nature of UOWC also necessitates efficient pointing-acquisitioning-and-tracking (PAT) mechanisms, especially for the collimated light beams propagating over a long range. Hence, accurate node location information has a critical importance for effective PAT mechanisms \cite{Saeed2018Underwater}. However, the estimated node locations may not refer to actual node positions either because of the localization errors or the random movements caused by the hostile underwater environment. These have negative impacts on the link reliability and performance as a result of the poor PAT efficiency. A potential solution for location uncertainty would be employing adaptive optics to provide a degree of robustness by adjusting the bandwidth to cover a wider spatial area around the estimated receiver location. 

Limited communication range and hostile underwater channel conditions also entail communicating over multiple hops in order to improve both connectivity and performance of UOWNs. In this respect, there is a dire need for analyzing the end-to-end (E2E) performance of multihop UOWCs under prominent relaying techniques. Moreover, multihop UOWCs require effective routing protocols that can tackle the range-beamwidth tradeoff and adversity of underwater environment. It is therefore of utmost importance to develop robust relaying and routing techniques which manipulate range-beamwidth tradeoff by means of adaptive optics, which is main focus of this paper. 

\subsection{Related Works}
Recent efforts on UOWC can be exemplified as follows: Arnon modeled three types of UOWC links: line-of-sight (LoS), modulating retroreflector, and  non-LoS (NLoS) \cite{Arnon10underwater}.  In \cite{Anous2018Vertical}, authors model and evaluate underwater visible light communication (UVLC) vertical links by dividing the UWON into layers and taking account of the inhomogeneous characteristics (e.g., refractive index, attenuation profiles, etc.) of each layer. Experimental evaluation of orthogonal frequency division multiplexing based UVLC is conducted in \cite{Hessien2018Experimental} and performance of automatic repeat request (ARQ) based UVLC system is analyzed in \cite{Shafique2017Performance}. Considering Poisson point process based spatial distribution of nodes, Vavoulas et. al. analyzed the $k$-connectivity of UOWNs under the assumption of omni-directional communications \cite{Vavoulas2014kconnectivity}. To take the directivity of UOWC into account, the $k$-connectivity analysis of UOWNs is readdressed in \cite{Saeed2018Connectivity} which is further extended to investigate impacts of connectivity on the localization performance \cite{Saeed2018performance}. 

Akhoundi et. al. introduced and investigated a potential adaptation of cellular code division multiple access (CDMA) for UOWNs \cite{Akhoundi2015cellular, Akhoundi2016cellular}. In \cite{Jamali2016performance}, authors characterized the performance of relay-assisted underwater optical CDMA system where multihop communication is realized by chip detect-and-forward method. Assuming identical error performance at each hop, Jamali et. al. consider the performance analysis of multihop UOWC using DF relaying \cite{Jamali2017multihop}. Since the commonly adopted Beer-Lambert UOWC channel model assumes that the scattered photons are totally lost, authors of \cite{Elamassie2018performance} modify this renown model to consider the received scattered photons by a single parameter which is a function of transceiver parameters and water type. Thereafter, they consider a dual-hop UVLC system and determine optimal relay placement to minimize the bit error rate (BER) for both DF and AF relaying.

\subsection{Main Contributions}
The main contributions of this paper can be summarized as follows:
\begin{itemize}
	\item  To model the displacement of the actual locations from the estimates, we consider an uncertainty disk whose radius is a design parameter that depends on the localization accuracy and/or mobility of nodes. Assuming nodes are equipped with adaptive optics, robust UOWC links are provisioned by deriving the divergence angles for scenarios with and without a PAT mechanism. Numerical results show that location accuracy and effective PAT mechanism are crucial to reach a desirable performance.

	\item  We investigate and compare two prominent relaying techniques; decode \& forward (DF) and optical amplify \& forward (AF), each of which is analyzed for important E2E performance metrics. For both relaying methods, closed form expressions are derived for various performance metrics, e.g., E2E-Rate, E2E bit error rate (BER), transmission power, amplifier gain, etc. For a given path, we also formulate and solve minimum total power consumption problems under both DF and AF relaying schemes.   

	\item In light of the adaptive divergence angles and E2E performance analysis, both centralized and distributed routing protocols are developed based on the availability of global or local network state information, respectively. The centralized techniques employ variations of shortest path algorithms which are tailored to achieve various objectives such as maximum achievable E2E-Rate, minimum E2E-BER, and minimum total power consumption. By manipulating the range-beamwidth tradeoff, we propose   \underline{Li}ght \underline{Pa}th \underline{R}outing (LiPaR); a distributed routing protocol that combines traveled distance progress and link reliability. Obtained results show that LiPaR can provide a superior performance compared to centralized schemes without a PAT mechanism.     
\end{itemize}

\subsection{Paper Organization}
The remainder of the paper is organized as follows: Section \ref{sec:sysmod} introduces the system model. Section \ref{sec:PAT} presents the uncertainty disk model and derives the adaptive divergence angles.   Section \ref{sec: relay} analyzes and compares DF and AF relaying techniques. Section \ref{sec:routing} develops the centralized and distributed routing protocols. Section \ref{sec:res} presents the numerical results. Finally, Section \ref{sec:conc} concludes the paper with a few remarks. 

\section{System Model}
\label{sec:sysmod}

\begin{figure}[!t]
\begin{center}
\includegraphics[width=1\columnwidth]{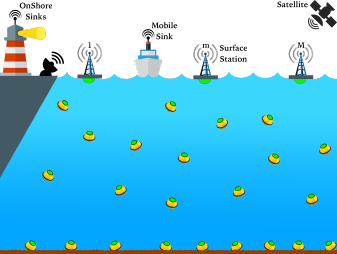}
\caption{Illustration of UOWNs with multiple sinks.}
\label{fig:uown}
\end{center}
\end{figure}

\begin{figure}[!t]
\begin{center}
\includegraphics[width=1\columnwidth]{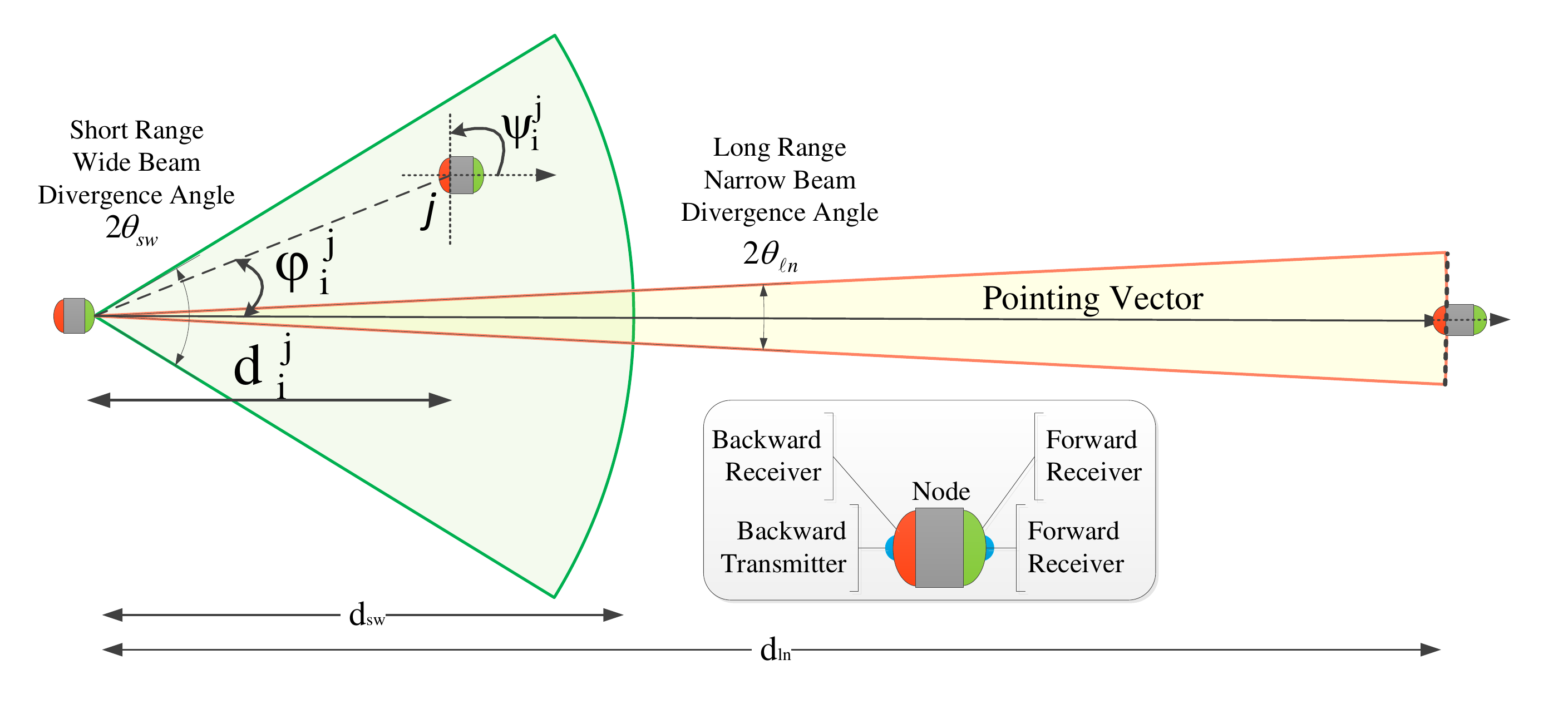}
\caption{Demonstration of the system model and range-beamwidth tradeoff.}
\label{fig:tradeoff}
\end{center}
\end{figure}

We consider a two-dimensional underwater optical wireless network (UOWN) which consists of a $M$ surface stations/sinks (SS) and $N$ nodes/sensors as shown in Fig. \ref{fig:uown}. The SSs are responsible for disseminating the data collected from sensors to mobile or onshore sinks. Nodes are equipped with two optical transceivers to enable bi-directional connections between sensors and SSs [c.f. Fig. \ref{fig:tradeoff}]. Transmitters are assumed to be capable of adapting their beam divergence angle $\theta_n$ within a certain range, i.e., $\theta_{\min} \leq \theta_n \leq \theta_{\max}$. The location estimates of the SS $m$ and node $n$ are denoted by $\boldsymbol{\ell}_m=[x_m, y_m], \: \forall m \in [1,M],$ and $\boldsymbol{\ell}_n=[x_n, y_n], \: \forall n \in [1,N],$ respectively. Likewise, actual locations of the SS $m$ and node $n$ are denoted by $\boldsymbol{\ell}_m^*=[x_m^*, y_m^*], \: \forall m \in [1,M],$ and $\boldsymbol{\ell}_n^*=[x_n^*, y_n^*], \: \forall n \in [1,N],$ respectively. In addition to $\boldsymbol{\ell}_n$, node $n$ is aware of $\boldsymbol{\ell}_m, \forall m$.

According to the Beer's law, aquatic medium can be characterized for wavelength $\lambda$ as a combination of absorption and scattering effects, i.e., $e(\lambda)= a(\lambda)+b(\lambda)$ where $a(\lambda)$, $b(\lambda)$ and $e(\lambda)$ are absorption, scattering and extinction coefficients respectively. These coefficients vary with water types (e.g., clear, harbor, turbid. etc.) and water depths (e.g., shallow, deep, etc.). The propagation loss factor of the LoS channel between node $i$ ($n_i$) and node $j$ ($n_j$) is defined as follows 
\begin{equation}
\label{eq:PL}
L_i^j=\exp \left \{ -e(\lambda) \frac{d_{ij}}{\cos(\varphi_i^j)} \right\},
\end{equation}
where $d_{ij}$ is the perpendicular distance and $\varphi_i^j$ is the angle between the receiver plane and the transmitter-receiver trajectory, i.e., the pointing vector of the transmitter. On the other hand, geometric loss of the LoS channel is given as \cite{Kahn97wireless}
\begin{equation}
\label{eq:cg}
g_{i}^{j}= \begin{cases} \frac{A_j}{d_{ij}^2} \frac{\cos(\varphi_i^j)}{2 \pi  [1-\cos(\theta_i)]} \xi(\psi_i^j)
, -\pi/2 \leq \varphi_i^j \leq \pi/2 \\
0, \hfill\text{otherwise}\end{cases},
\end{equation}
where $A_j$ is the receiver aperture area of $n_j$, $\theta_i^j$ is the half-beamwidth divergence angle of $n_i$\footnote{$\theta_i^j$ is measured at the point where the light intensity drops to $1/e$ of its peak.}, and $\xi(\psi_i^j)$ is the concentrator gain, that is defined as
\begin{equation}
\label{eq:LOSgain}
\xi(\psi_i^j)=\begin{cases} \frac{\iota^2}{\sin^2(\Psi_j)}, 0 \leq \psi_i^j \leq \Psi_j \\
0, \hfill \psi_i^j> \Psi_j \end{cases},
\end{equation}
 $\psi_i^j$ is the angle of incidence w.r.t. the receiver axis, $\Psi_j$ is the concentrator field-of-view (FoV)\footnote{$\Psi_j$ can be $\pi/2$ and down to $\pi/6$ for hemisphere and parabolic concentrators, respectively.}, and $\iota$ is the internal refractive index.  

As shown in Fig. \ref{fig:tradeoff}, UOWC channel characteristics pose a fundamental tradeoff between communication range and divergence angle (i.e., beamwidth). That is, a wide divergence angle can reach nearby nodes whereas a narrow divergence angle is able to reach distant destinations. Therefore, manipulation of this tradeoff has significant impacts on several performance metrics such as the degree of connectivity, distance progress for the next hop, routing efficiency, and E2E performance. For instance, a wide divergence angle provides more path diversity at the expense of a higher number of hops which increases energy consumption and E2E delay. On the other hand, a narrow divergence angle can deliver a desirable performance over long distances, which requires precise and agile PAT mechanisms to keep the transceivers aligned over long ranges. Throughout the paper, this tradeoff is to be discussed in the E2E performance analysis of different relaying techniques and the design of the routing protocols.       

\section{Adaptive Beamwidth and PAT Mechanism}
\label{sec:PAT}
For a reliable and well-connected transmission, the inherent directivity of UOWC requires perfectly aligned point-to-point links between the transceivers. First and foremost information needed for a proper alignment is node coordinates that can be obtained either by pure optical \cite{Saeed2018Robust} or hybrid acoustic-optical localization algorithms \cite{Saeed2017Energy}. Nonetheless, localization errors introduce uncertainty over the location estimates, which may be further exacerbated by the random movements of nodes due to the hostile underwater conditions. Thus, the accuracy of location estimates plays an essential role in the E2E performance of the multihop communication. One solution to this problem is adaptively changing the properties of the light beam. Indeed, misalignment caused by the location uncertainty can be alleviated by adapting the divergence angle to cover a larger spatial region at the cost of degraded performance. In the remainder, we therefore probe an initial investigation into the robust divergence angles and its impact on the E2E performance by considering the following three different cases.

\begin{figure}[!t]
\begin{center}
\includegraphics[width=1\columnwidth]{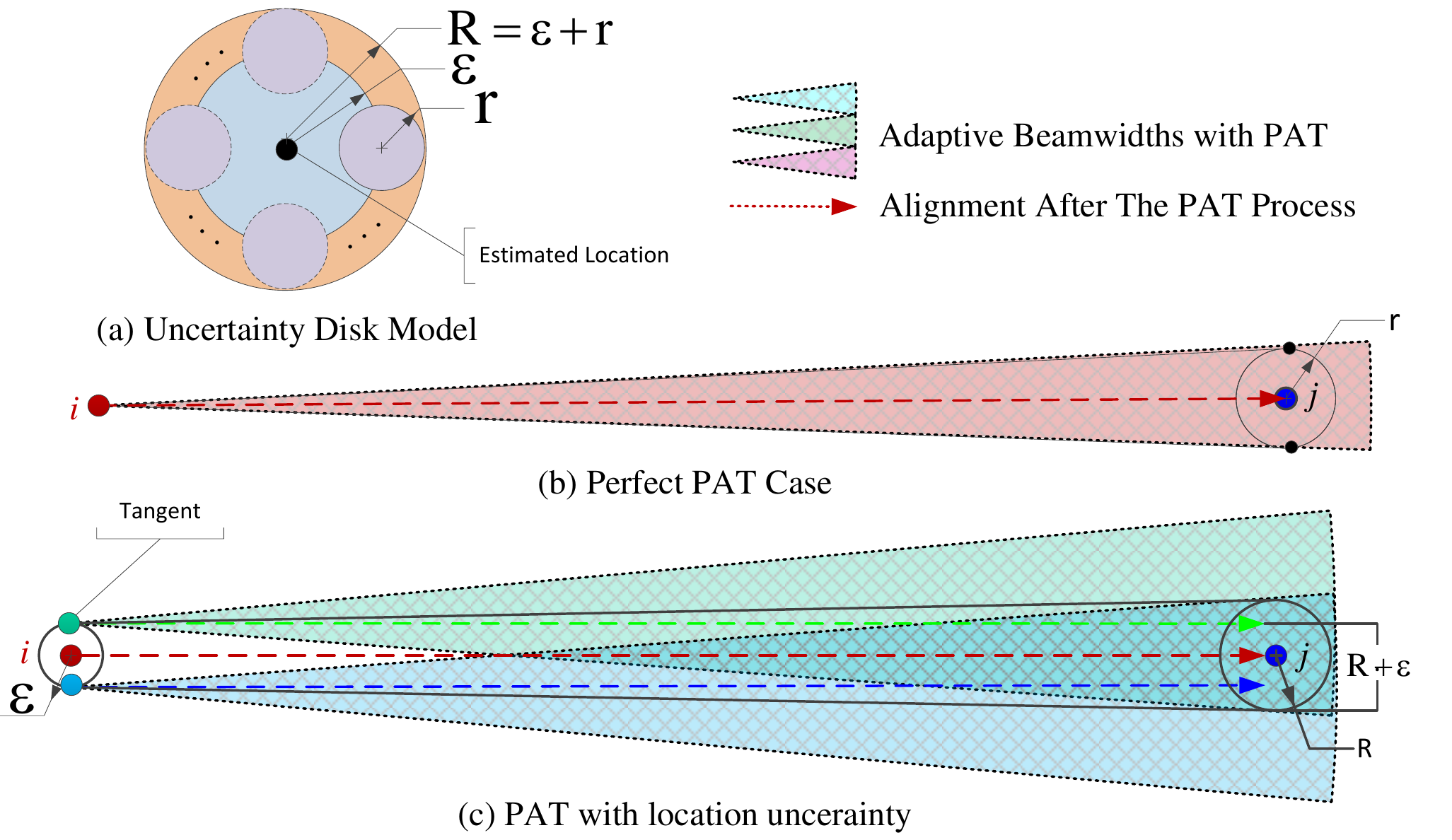}
\caption{Adaptive beamwidths for PAT mechanisms.}
\label{fig:pat1}
\end{center}
\end{figure}

\subsection{Divergence Angles for PAT with Perfect Location Estimates}
In the ideal case, estimated and actual locations are the same (i.e.,  $\epsilon=0$), the pointing vector is aligned to $n_j$, and divergence angle is adjusted to exactly cover the node frame, which is shown by plaid sectors in Fig. \ref{fig:pat1}.b. Accordingly, the full divergence angle of the perfect PAT scenario is given by
\begin{equation}
\label{eq:theta_pp}
\theta_{ij}^{pp}(r)= \max\left(\theta_{min}, \arcsin \left( \frac{r}{d_{ij}}\right)\right),
\end{equation}
which follows from the law of sines and the fact that radius is perpendicular to the tangent points. 

\subsection{Divergence Angles for PAT under Location Uncertainty}
The location uncertainty can be caused by estimation errors of localization method and/or the random movements of the transceivers, which results in a radial displacement of the actual location from the estimated location. As shown in Fig. \ref{fig:pat1}.a, we model node locations by an uncertainty disk centered at the location estimates and has a radius of $R=r+\epsilon$ where $r$ is the radius of the node frame and $\epsilon$ is the uncertainty metric. Under the location uncertainty, the divergence angle must be calculated based on the worst case scenarios where the transmitter is located at one of the tangents on the uncertainty disk [c.f. Fig. \ref{fig:pat1}.c]. In this case, the divergence angle is twice of the perfect angle covering the uncertainty disk, i.e, 
\begin{equation}
\label{eq:theta_ip}
\theta_{ij}^{ip}(\epsilon,r)= \max\left(\theta_{min}, \arcsin \left( \frac{2\epsilon+r}{d_{ij}}\right)\right),
\end{equation}
which can be defined as a robust divergence angle as it assures the worst case beamwidth to establish a link between the transceivers. Noting that this is the first step of the PAT procedure, i.e., pointing, acquisitioning and tracking are the next steps to keep nodes aligned via feedbacks, which results in Fig. \ref{fig:pat1}.b. Although developing a PAT mechanism is out of our scope, our purpose is to show its critical role in the multihop UOWN performance.   

\begin{figure}[!t]
\begin{center}
\includegraphics[width=1\columnwidth]{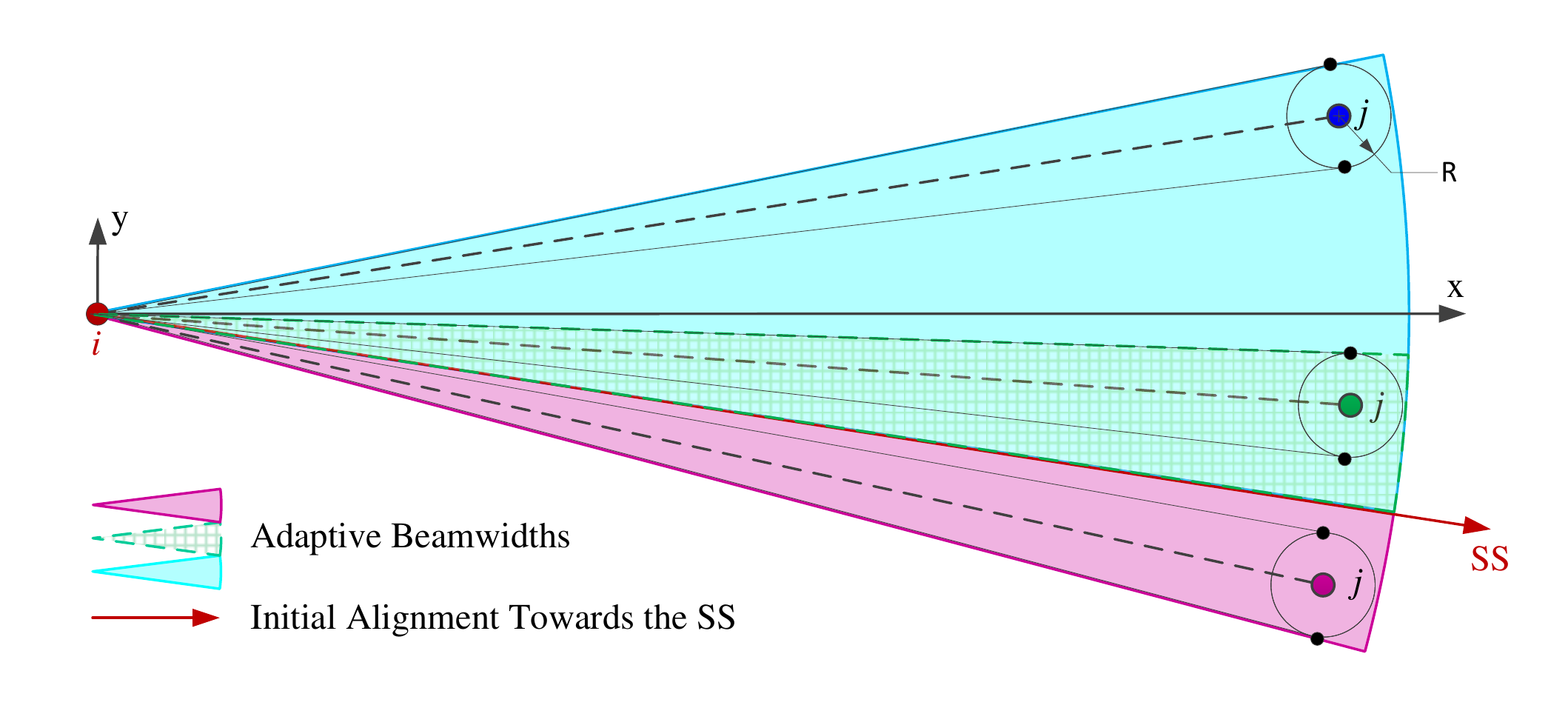}
\caption{Adaptive half-beamwidths under the receiver location uncertainty.}
\label{fig:pat2}
\end{center}
\end{figure}

\begin{figure}[!t]
\begin{center}
\includegraphics[width=1\columnwidth]{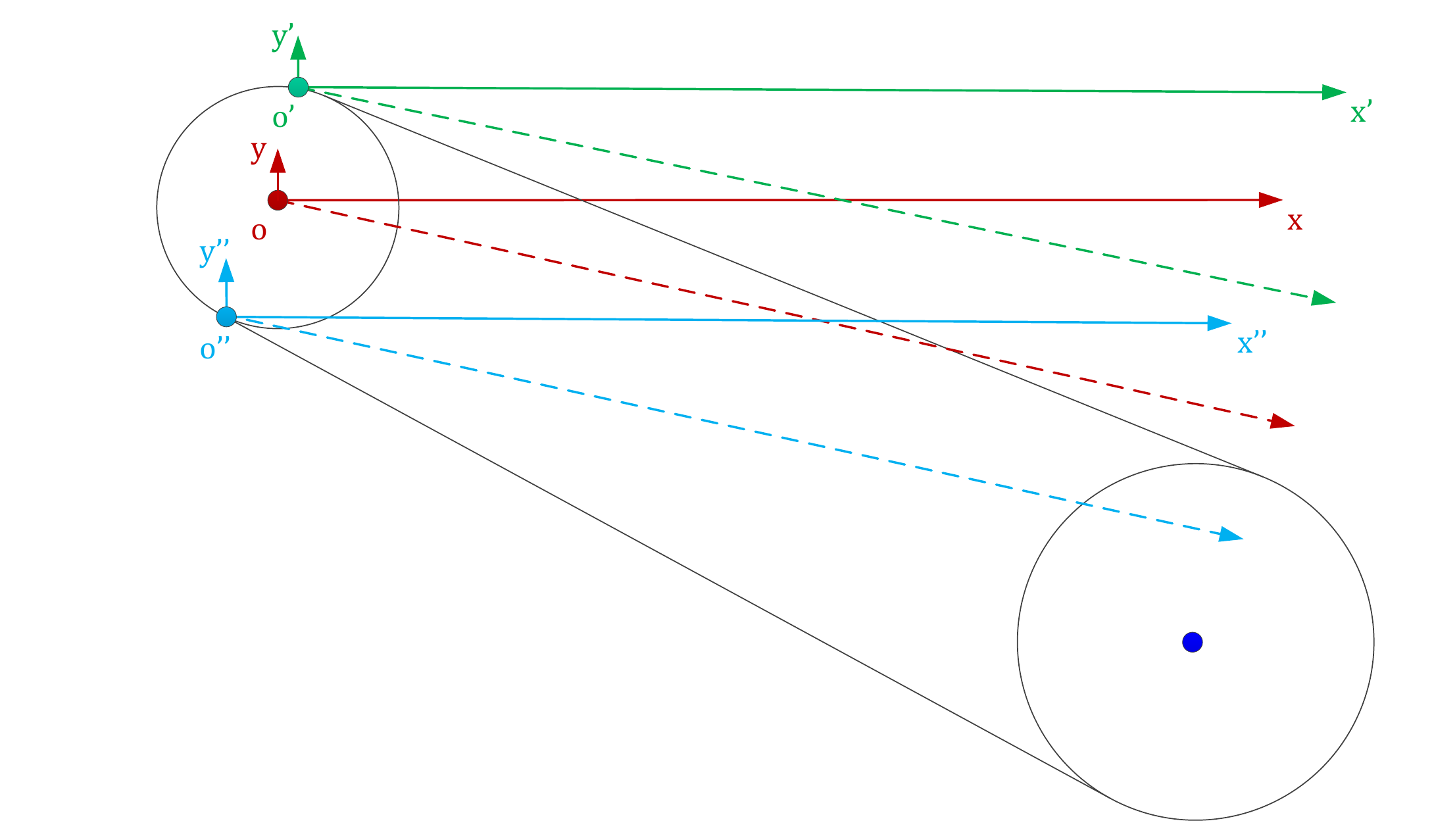}
\caption{Illustration of adaptive beamwidth calculations under the transceiver location uncertainty.}
\label{fig:pat3}
\end{center}
\end{figure}

\subsection{Divergence Angles under the Absence of PAT}
In the absence of the PAT mechanism, we assume that nodes keep their body frame directed towards the location of the closest SS to align transmitter's pointing vector with the SS receiver. Although pointing to a certain location is not possible, nodes can still reach nodes within a certain angle around the pointing vector by adjusting the beamwidth. 

For the sake of a clear presentation, let us first consider the location uncertainty in the receiver node only [c.f. Fig. \ref{fig:pat2}], where the origin of the local Cartesian coordinate system is set to the estimated transmitter location, i.e., $\boldsymbol{o}=\boldsymbol{\ell}_i=\boldsymbol{\ell}_i^*$. In this local coordinate system, the locations of $n_j$ and the destination SS $m$ (SS$_m$) are given by $\boldsymbol{\tilde{\ell}}_j=[x_j-x_i, y_j-y_i]$ and $\boldsymbol{\tilde{\ell}}_m=[x_m-x_i, y_m-y_i]$, respectively. Hence, the angles between the x-axis and vectors pointing $n_j$ ($\overrightarrow{\boldsymbol{\ell}_i \boldsymbol{\ell}_j}$) and the SS$_m$ ($\overrightarrow{\boldsymbol{\ell}_i \boldsymbol{\ell}_m}$) are given by 
\begin{equation}
\label{eq:phi_gamma}
\varphi_i^j=\arctan\left(\frac{y_j-y_i}{x_j-x_i}\right) \text{ and }\phi_i^{s}=\arctan\left(\frac{y_m-y_i}{x_m-x_i}\right),
\end{equation}
respectively. Accordingly, the divergence angle that is centered around $\overrightarrow{\boldsymbol{\ell}_i \boldsymbol{\ell}_m}$ covering the uncertainty disk is given by \eqref{eq:theta_unc_j} where $\sgn(\cdot)$ is the signum function.        
\begin{figure*}
\begin{equation}
\label{eq:theta_unc_j}
\theta_{ij}^o=
\begin{cases}
 \theta_{ij}^{ip}(\epsilon,r) &, \text{if } \phi_i^{s}=\varphi_i^j \\
2 \left \vert \phi_i^{s} -\varphi_i^j \right\vert +\frac{\theta_{ij}^{pp}(R)}{2} &, \text{if }\sgn(\phi_i^{s})=\sgn(\varphi_i^j) \bigwedge \phi_i^{s} \neq \varphi_i^j  \\
 2 \left \vert \varphi_i^j - \phi_i^{s} \right\vert + \frac{\theta_{ij}^{pp}(R)}{2}   &, \text{if }\sgn(\phi_i^{s}) \neq \sgn(\varphi_i^j) \\
2 \max \left(  \left \vert \phi_i^{s} - \left( \varphi_i^j + \frac{\theta_{ij}^{pp}(R)}{2}\right) \right\vert, \left \vert \phi_i^{s} - \left( \varphi_i^j - \frac{\theta_{ij}^{pp}(R)}{2} \right) \right\vert  \right) &, \text{if } \varphi_i^j+\frac{\theta_{ij}^{pp}(R)}{2} \geq \phi_i^{s} \geq \varphi_i^j-\frac{\theta_{ij}^{pp}(R)}{2}
\end{cases}
\end{equation}
\hrule
\end{figure*}
Fig. \ref{fig:pat2} illustrates the half divergence angle of the first and second cases in \eqref{eq:theta_unc_j} by green/red and blue colors, respectively. The last case is the scenario where $\overrightarrow{\boldsymbol{\ell}_i \boldsymbol{\ell}_m}$ passes through the uncertainty disk. Similar to the previous subsection, we consider the wost case scenario when there is uncertainty about both receiver and transmitter locations. Therefore, we consider two additional local coordinates originated at tangent locations [c.f. Fig. \ref{fig:pat3}] which are given by
\begin{align}  
\boldsymbol{o'}&=\left[x_i+\epsilon \cos \left(\gamma_i^s+\frac{\pi}{2}\right), y_i+\epsilon \sin \left(\gamma_i^s+\frac{\pi}{2}\right)\right]\\
\boldsymbol{o''}&=\left[x_i+\epsilon \cos \left(\gamma_i^s-\frac{\pi}{2}\right), y_i+\epsilon \sin \left(\gamma_i^s-\frac{\pi}{2}\right)\right].
\end{align}  
Noting that $\gamma_i^s$ is the same in all local coordinates, $\varphi_i^j$ for $\boldsymbol{o'}$ ($\varphi_{ij}^{'}$) and $\boldsymbol{o''}$ ($\varphi_{ij}^{''}$) should be obtained as in \eqref{eq:phi_gamma}. By substituting $\varphi_{ij}^{'}$ and $\varphi_{ij}^{''}$ into \eqref{eq:theta_unc_j}, one can obtain divergence angles for transmitters located at tangent points $\boldsymbol{o'}$ ($\theta_{ij}^{'}$) and $\boldsymbol{o''}$ ($\theta_{ij}^{''}$), respectively. In case of the location uncertainty at both sides, the divergence angle is then given by 
\begin{equation}
\label{eq:theta_no_pat}
\theta_i^j=\max\left(\theta_{min},\max\left(\theta_{ij}^{o},\theta_{ij}^{'},\theta_{ij}^{''}\right)\right).
\end{equation}

\section{Performance Analysis of Relaying Techniques}
\label{sec: relay}
In this section, we consider the E2E performance analysis of potential relaying techniques for an arbitrary multi-hop path from a source node to one of the sink nodes. Such a path is defined as an ordered set of nodes, i.e., $\mathcal{H}_{o \rightsquigarrow s} = \{ h \vert  0 \leq h \leq H\}$ where $H=|\mathcal{H}_{o \rightsquigarrow s}|$ is the number of hops and the first (last) element represents the source (sink). 

\subsection{Decode \& Forward Relaying}
\label{sec:DF}
In DF transmission, the received optical signal at each hop is converted into electrical signal, then decoded, and finally re-encoded before retransmission for the next hop. Thus, the DF relaying requires high speed data converters, decoders, and encoders to sustain the achievable data rates in the order of Gbps. For an arbitrary path, received power at node $h$ is given as 
\begin{equation}
\label{eq:Pr_LoS}
P_h^r= P_{h-1}^t \eta_{h-1}^t \eta_h^r G_{h-1}^h, \: \forall h \in [1, H],
\end{equation}
where $P_{h-1}^t $ is the average optical transmitter power of previous node, $\eta_{h-1}^t$ is the transmitter efficiency of node $h-1$, $\eta_h^r$ is the receiver efficiency of node $h$, and $G_{h-1}^h= L_{h-1}^h g_{h-1}^h$ is the composite channel gain.

The most common detection technique for OWC is intensity-modulation/direct-detection (IM/DD) with on-off keying (OOK). Assuming that photon arrivals follows a Poisson process, photon arrival rate of node $h$ is given as \cite{Arnon09nonLOS}
\begin{equation}
\label{eq:photon_count}
f_{h}(p)=\frac{P_h \eta_h^d \lambda}{R_{h-1}^h T \hslash c}, \: \forall h \in [1, H],
\end{equation}
where $P_h$ is the observed power at node $h$, $\eta_h^d$ is the detector counting efficiency, $R_{h-1}^h$ is the transmission rate for the hop $(h-1,h)$, $T$ is pulse duration, $\hslash$ is Planck's constant, and $c$ is the underwater speed of light. Hence, the photon counts when binary '0' transmitted is given by $p_{h}^0\triangleq f_{h}(P_n)$ where $P_n=P_{dc}+P_{bg}$ is the total noise power, $P_{dc}$ is the additive noise power due to dark counts and $P_{bg}$ is the background illumination noise power. Similarly, the photon counts when binary '1' transmitted is given by $p_{h}^1 \triangleq f_{h}(P_h^r+P_n)$. Assuming a large number of photon reception, the Poisson distribution can be approximated by a Gaussian distribution as per the Central Limit Theorem. For a given data rate, $\bar{R}$ , the BER of a single hop is given by \cite{Arnon10underwater}
\begin{equation}
\label{eq:BER}
\mathcal{P}_{h-1}^h=\frac{1}{2}\erfc \left(\sqrt{\frac{T}{2}} \left[ \sqrt{p_{h}^1} - \sqrt{p_{h}^0}\right] \right)
\end{equation}
where $\erfc(\cdot)$ is the complementary error function. For a certain BER threshold, $\bar{\mathcal{P}}_{i,j}^e$, data rate of the hop $(h-1,h)$ is derived from (\ref{eq:BER}) as
\begin{equation}
\label{eq:rate}
R_{h-1}^h=\frac{\eta_h^d \lambda}{2 \hslash c} \left[\frac{\sqrt{P_h^r+P_n}-\sqrt{P_n}}{\erfc^{-1} (2 \bar{\mathcal{P}}_{h-1}^h)}\right]^2,
\end{equation}
which is inversely proportional to the BER. 
 Following from \eqref{eq:Pr_LoS}-\eqref{eq:rate}, minimum transmit power to ensure a predetermined rate $\bar{R}_{h-1}^h$ and BER $\bar{\mathcal{P}}_{h-1,h}^e$ is given by
\begin{equation}
\label{eq:DF_Pt}
P_{h-1}^t =\frac{a^2+2a\sqrt{P_n
}}{G_{h-1}^h \lambda \eta_{h-1}^t \eta_h^r  \eta_h^d}
\end{equation}
where $a=\erfc^{-1} (2 \bar{\mathcal{P}}_{h-1}^h) \sqrt{\frac{2\bar{R}_{h-1}^h \hslash c }{\eta_h^d \lambda}}$. For a given error and data rate, communication range between two generic nodes ($i,j$) is given by
\begin{align}
\label{eq:DLos}
D_i^j&=\frac{2}{e(\lambda)}W_0 \left( \frac{e(\lambda)}{2\sqrt{c}\cos(\phi_i^j)} \right)
\end{align}
where  $W_0(\cdot)$ is the principal branch of product logarithm function, $c=\frac{2 b \pi [1-\cos(\theta_i)]}{A_j \cos(\phi_i^j) \xi(\psi_i^j)}$ and $b=\frac{a^2+2a\sqrt{P_n}}{P_{h-1}^t\lambda \eta_{h-1}^t \eta_h^r  \eta_h^d}$. We refer interested readers to Appendix \ref{app:range} for derivation details.

Before proceeding into the E2E performance analysis, it is necessary to discuss single-hop performance in light of Section \ref{sec:PAT} and the above derivations. For a given transmit power and communication range, increasing the divergence angle deteriorate the $\mathcal{P}_{h-1,h}^e$ and $R_{h-1}^h$, which are inversely proportional as per \eqref{eq:BER} and \eqref{eq:rate}. On the contrary, communication range decreases by increasing (decreasing) $R_{h-1}^h$ ($\mathcal{P}_{h-1,h}^e$) and decreasing (increasing) the transmit power (divergence angle). Noting the coupled relation of the divergence angle with all these metrics, single-hop performance analysis clearly shows the need for effective PAT mechanisms to obtain a superior E2E performance, which is analyzed next. 

Denoting $\mathcal{X}_h$ as the Bernoulli random variable which represent the erroneous decision of node $h$, total number of incorrect decision made along the path is given as $\mathcal{X}=\sum_{h=1}^{H} \mathcal{X}_h$. Assuming $\mathcal{X}_h$'s are independent but non-identically distributed, $\mathcal{X}$ is a Poisson-Binomial random variable. Since a transmitted bit is received correctly at the sink if the number of erroneous detection is even, E2E BER is derived as  
\begin{align}
\label{eq:PE2E}
\mathcal{P}^{DF}_{E2E}&=\sum_{j \in \mathcal{A}} \sum_{ \mathcal{B} \in \mathcal{F}_{j}}  \prod_{k \in \mathcal{B}}\mathcal{P}_{k-1}^k \prod_{l\in \mathcal{B}^c} \left(1-\mathcal{P}_{l-1}^{l}\right)
\end{align}
where $\mathcal{A}=\{1,3,\ldots, H\}$ is the set of odd numbers and $\mathcal{F}_{j}$ is the set of all subsets of $j$ integers that can be selected from $\mathcal{A}$. $\mathcal{P}_{E2E}$ can be expeditiously calculated from polynomial coefficients of the probability generating function of $\mathcal{X}$ in $O(A \log A)$ where $A$ is the cardinality of  $\mathcal{A}$ \cite{fernandez2010closed}. Notice that $\mathcal{X}$ reduces to a Binomial variable if all hops are assumed to be identical as in \cite{Jamali2017multihop} , which is hardly the case in practice. 

The achievable E2E rate is determined by the minimum of the data rates along the path, i.e.,
\begin{equation}
\label{eq:E2Erate}
R_{E2E}^{DF}= \underset{1 \leq h \leq H}{\min} \left( R_{h-1}^h \right),
\end{equation}
which implies that a predetermined E2E data rate $\bar{R}_{E2E}$ requires $R_{h-1}^h \geq \bar{R}_{E2E}, \forall h \in [1,H]$. Accordingly, $\vect{\mathrm{P}_1}$ formulates the optimization problem which minimizes the total transmission powers while ensuring a target E2E BER rate, $\bar{\mathcal{P}}_{E2E}$, and E2E data rate by setting $R_{h-1}^h = \bar{R}_{E2E}, \forall h$.   
\begin{equation}
\hspace*{0pt}
 \begin{aligned}
 \label{eq:P1}
 & \hspace*{0pt} \vect{\mathrm{P}_1}:\: \: \underset{\vect{\mathcal{P}}}{\min}
& & \hspace*{3 pt} \vect{1}^{T}\vect{P_t} \\
& \hspace*{0pt} \mbox{$\mathrm{P}_1^1$: }\hspace*{5pt} \text{s.t.}
&&  \log \left(\mathcal{P}_{E2E} \right) \leq \log \left(\bar{\mathcal{P}}_{E2E}\right)\\
 &
  \hspace*{0 pt}\mbox{$\mathrm{P}_1^2$: } & & \vect{0}  \preceq \vect{\mathcal{P}} \preceq \vect{0.5}
  \end{aligned}
\end{equation}
where $\vect{\mathcal{P}}$ is the BER vector of the path, $\vect{1}^{T}$ is the transpose of vector of ones, and $\preceq$ denotes the pairwise inequality for vectors. Based on the mild assumption of ensuring each hop has a BER no more than $0.5$, $\vect{\mathrm{P}_1}$ can be shown to be a convex problem using the log-concavity of the Poisson-Binomial distribution. We refer interested readers to Appendix \ref{app:p1} for a formal convexity analysis of $\vect{\mathrm{P}_1}$.

\begin{figure}[!t]
\begin{center}
\includegraphics[width=1\columnwidth]{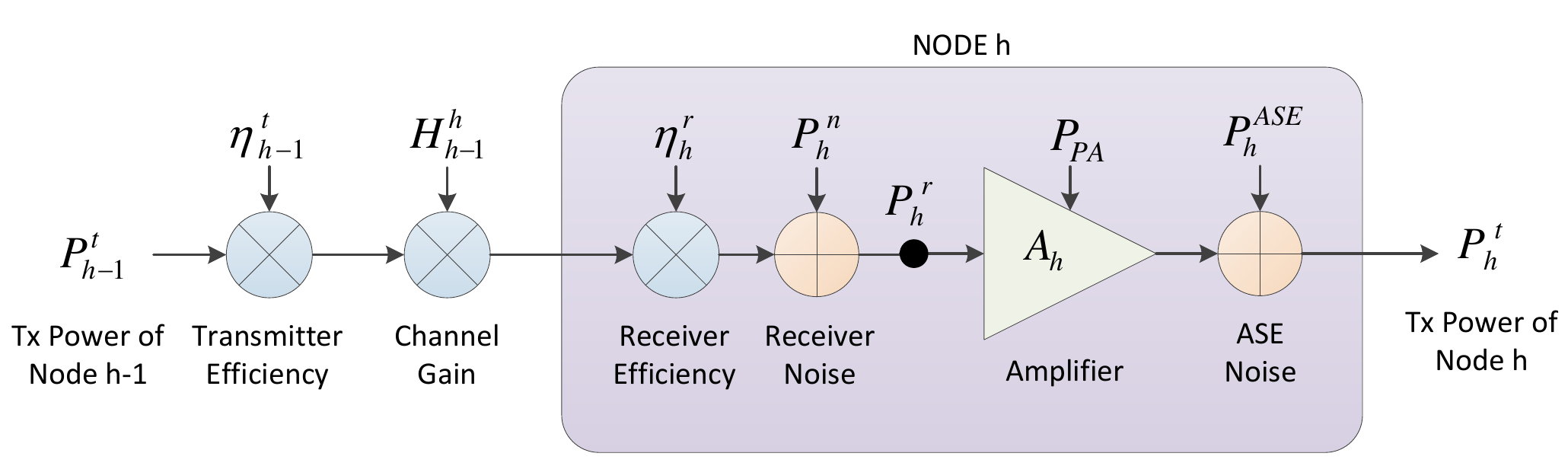}
\caption{Receiver diagram for the optical AF and RF relays.}
\label{fig:AF}
\end{center}
\end{figure}

\subsection{Optical Amplify \& Forward Relaying}
\label{sec:AF}
Although DF greatly improves the E2E performance by limiting the background noise propagation, it introduces extra power consumption and signal processing delay. Furthermore, synchronization and clock recovery are additional challenges to be tackled in Gbps links. Alternatively, the AF relaying executes optical-to-electrical (OEO) conversion at each node, which amplifies the received signal electrically and then retransmits the amplified signal for the next hop. The main drawback of the AF transmission is propagation of noise added at each node, which is amplified and accumulated through the path. As a remedy to costly OEO conversion, all optical AF relaying has the advantages of realizing the high speed transmissions without the need for OEO conversion and sophisticated optoelectronic devices. As illustrated in Fig. \ref{fig:AF}, amplified and transmitted power of all-optical AF scheme can be modeled for intermediate nodes $1 \leq h \leq H-1$  as follows
\begin{align}
\label{eq:Rx_model}
P_h^r&=P_{h-1}^t \eta_{h-1}^t \eta_{h}^r G_{h-1}^{h} +P_n, \\
\label{eq:Tx_model}
P_h^t&= A_h P_h^r+P_h^{a},
\end{align}
where $P_n=P_{bg}+P_{dc}$ is the local noise at the receiver, $A_h$ is the amplifier gain, $P_{a}=N_o^h B$ is the amplified spontaneous emission (ASE) noise that is modeled as an additive zero-mean white Gaussian, $N_o^h= \hslash f_0 \left( G_{h-1}^{h} -1 \right) n_{sp}$ is the power spectral density per polarization, $f_0$ is the frequency, $n_{sp}$ is spontaneous emission parameter of the amplifier, and $B$ is the amplifier bandwidth \cite{agrawal2012fiber}. By assuming the independence of signals, channel gains, and noises, amplified and transmitted powers at node $h$ can be respectively written in the following generic form 
\begin{align}
\label{eq:Rx@h}
\nonumber P_h^r&=P_t^0 \prod_{i=1}^{h-1}A_i \prod_{j=0}^{h-1}\eta_j^t \prod_{k=1}^{h} \eta_k^r  \prod_{l=1}^{h} G_{l-1}^{l} \\
\nonumber  &+ \sum_{i=1}^{h} P_n   \prod_{j=i}^{h-1}A_j \prod_{k=i}^{h-1}\eta_k^t \prod_{l=i+1}^{h} \eta_l^r  \prod_{m=i+1}^{h} G_{m-1}^{m} \\
 &+ \sum_{i=1}^{h-1} P_{a}   \prod_{j=i+1}^{h-1}A_j \prod_{k=i}^{h-1}\eta_k^t \prod_{l=i+1}^{h} \eta_l^r  \prod_{m=i+1}^{h} G_{m-1}^{m}, 
 \end{align}
\begin{align}
\label{eq:Tx@h}
\nonumber P_h^t&= P_t^0 \prod_{i=1}^{h}A_i \prod_{j=0}^{h-1}\eta_j^t \prod_{k=1}^{h} \eta_k^r  \prod_{l=1}^{h} G_{l-1}^{l} \\
\nonumber  &+ \sum_{i=1}^{h} P_n   \prod_{j=i}^{h}A_j \prod_{k=i}^{h-1}\eta_k^t \prod_{l=i+1}^{h} \eta_l^r  \prod_{m=i+1}^{h} G_{m-1}^{m}  \\
 &+ \sum_{i=1}^{h} P_{a}   \prod_{j=i+1}^{h}A_j \prod_{k=i}^{h-1}\eta_k^t \prod_{l=i+1}^{h} \eta_l^r  \prod_{m=i+1}^{h} G_{m-1}^{m}, 
\end{align}
which follows from the recursion of (\ref{eq:Rx_model}) and (\ref{eq:Tx_model}). 
Because of the propagating noise along the path, there is no a unique way of ensuring a target E2E performance. However, a practical and tractable means of analyzing the E2E performance is fixing the transmission power at each hop.  Noting that the combination of background light and dark current noise is more dominant than the ASE noise \cite{karp1988optical}, the amplifier gain can be calculated as 
\begin{equation}
\label{eq:gain}
A_h=\frac{P_t}{G_{h-1}^{h} \eta_{h-1}^t \eta_{h}^r+P_n}
\end{equation}
which is obtained by substituting \eqref{eq:Rx_model} into \eqref{eq:Tx_model}, equalizing \eqref{eq:Tx_model} to a fixed transmission power $P_t$, and then solving for $A_h$. Let us denote the transmitted signal-to-noise-ratio (SNR) at the source node as $\gamma=\frac{P_t}{P_n}$ and the received SNR in relay node $h$ as $\gamma_h$. By substituting \eqref{eq:gain} into \eqref{eq:Tx@h} and neglecting the ASE related last term of \eqref{eq:Tx@h}, the received SNR at the sink node can be obtained after some algebraic manipulations as 
\begin{equation}
\gamma_H=\frac{1}{\prod_{h=1}^{H}\left(1+\frac{1}{\gamma_h}\right)-1}
\end{equation}
where $\gamma_h=G_{h-1}^{h} \eta_{h-1}^t \eta_{h}^r \gamma$. Accordingly, the E2E-BER of optical AF relaying over an arbitrary route is given by 
\begin{equation}
\label{eq:AF_BER}
\mathcal{P}_{E2E}^{AF}=\frac{1}{2}\erfc \left(\sqrt{\frac{T p_{h}^0}{2}} \left[\sqrt{\frac{\prod_{h=1}^{H}\left(1+\gamma_h^{-1}\right)}{\prod_{h=1}^{H}\left(1+\gamma_h^{-1}\right)-1}} - 1\right] \right)
\end{equation}
which is derived by rewriting \eqref{eq:BER} for $\gamma_H$. For a given E2E-BER target $\bar{\mathcal{P}}_{E2E}$, the achievable E2E data rate of AF scheme is derived as  
\begin{equation}
\label{eq:AFrate}
R_{E2E}^{AF}=\frac{P_n \eta_h^d \lambda\left[\sqrt{\frac{\prod_{h=1}^{H}\left(1+\gamma_h^{-1}\right)}{\prod_{h=1}^{H}\left(1+\gamma_h^{-1}\right)-1}} - 1\right]^2}{2 \hslash c \left[ \erfc^{-1} \left(2 \bar{\mathcal{P}}_{E2E}\right)\right]^2},
\end{equation}
which follows from \eqref{eq:photon_count} and \eqref{eq:AF_BER}. Accordingly, the minimum total transmit power, that ensures a given E2E data and error rate pair, can be obtained by solving the following problem
\begin{equation}
\label{eq:P2}
\vect{\mathrm{P}_2}: \min_{0\leq P_t\leq \hat{P}_t} H P_t \:\: s.t. \:\: R_{E2E}^{AF} \geq \bar{R}_{E2E}
\end{equation}
where $\hat{P}_t$ is the maximum transmit power. Since $\gamma_H$ is a monotonically increasing function of $P_t$, the optimal transmit power $P_t^\star$ can be easily obtained by standard line search methods. Accordingly, optimal amplifier gains, $A_h^\star, \: \forall h$, that assures minimum E2E energy cost can be calculated by substituting $P_t^\star$ into \eqref{eq:gain}.


\section{Centralized and Distributed Routing Schemes}
\label{sec:routing}
Depending on the available network information, routing protocols can be designed either in distributed or centralized fashion. In this sense, distributed solutions are suitable for scenarios where nodes have local information about the network state of its neighborhood. On the other hand, centralized routing relies upon the availability of the global network topology and thus yields a superior E2E performance. However, it is worth mentioning that collecting a global network state information may yield extra communication overhead and energy cost. Therefore, the remainder of this section addresses centralized and distributed UOWN routing techniques. 

\subsection{Centralized Routing Schemes}

Let us represent the network by a graph $\mathcal{G}\left(\mathcal{V}, \mathcal{E}, \Omega \right)$ where $\mathcal{V}$ is the set of sensor nodes, $\mathcal{E}$ is the set of edges, and $\Omega \in \mathbb{R}^{N \times N}$ is the edge weight matrix that can be designed to achieve different routing objectives. In this respect, the remainder of this section considers three primary routing objectives: minimum E2E-BER, maximum E2E-Rate, and minimal total power consumption.  

\subsubsection{Minimum E2E-BER Routing}
\label{sec:bsr}

The main goal of this routing scheme is to find a path that minimizes the E2E-BER, $\mathcal{P}^{DF}_{E2E}$, while ensuring a predetermined E2E data rate, $\bar{R}_{E2E}$. As previously shown in \eqref{eq:E2Erate}, we satisfy this constraint by setting the data rate of each hop to $\bar{R}_{E2E}$. By neglecting the fortunate events of corrections at even number of errors, we make the following mild assumption for the DF relaying; the sink can correctly detect a bit if-and-only-if it is successfully detected at each hop.  

Thus, we equivalently consider the maximization of the E2E bit success rate (BSR), $\mathrm{BSR} \triangleq \prod_{h=1}^{H} \left(1-\mathcal{P}_{h-1}^h\right)$, instead of minimizing the $\mathcal{P}^{DF}_{E2E}$. Since the shortest path algorithms are only suitable to additive costs, BSR maximization can be transformed from the multiplication into a summation as follows
\begin{align}
\label{eq:BERroute}
\underset{\forall (o \rightsquigarrow s)}{\max} (\mathrm{BSR}) &= \underset{\forall (o \rightsquigarrow s)}{\max} \log \left(\prod_{h=1}^{H} \left(1-\mathcal{P}_{h-1}^h\right) \right) \\
&= \underset{\forall (o \rightsquigarrow s)}{\min}  \left( \sum_{h=1}^{H}  -\log\left(1-\mathcal{P}_{h-1}^h\right) \right)
\end{align}
where $o \rightsquigarrow s$ denotes an arbitrary path. By putting the product into an additive form, the maximum BSR route can be calculated by \textit{Dijkstra's shortest path} (DSP) algorithm with a time complexity of $\mathcal{O}\left( \vert \mathcal{V} \vert^2 \right)$. Accordingly, edge weights is set to the non-negative values of $\Omega_i^j= - \log  ( 1-\mathcal{P}_{i}^{j})$ where $\mathcal{P}_{i}^{j}$ is the BER between $n_i$ and $n_j$. Once the route is calculated, $\mathcal{P}^{DF}_{E2E}$ can be computed by substituting $\bar{R}_{E2E}$ into \eqref{eq:PE2E}.

In the case of AF relaying, optimal route must account for the noise propagation along the path. Following from \eqref{eq:AF_BER} and \eqref{eq:AFrate}, both E2E data rate and BER can be minimized by maximizing the SNR at the destination node, $\gamma_H$. Based on the same rationale in \eqref{eq:BERroute}, maximizing the SNR can be transformed into an additive form as follows  
\begin{align}
\label{eq:AFBERroute}
\underset{\forall (o \rightsquigarrow s)}{\max} \gamma_H &= \underset{\forall (o \rightsquigarrow s)}{\min} \prod_{h=1}^{H}\left(1+\gamma_h^{-1}\right) \\
&=\underset{\forall (o \rightsquigarrow s)}{\min} \log \left(\prod_{h=1}^{H}\left(1+\gamma_h^{-1}\right) \right) \\
\label{eq:AFBERroute3}
&= \underset{\forall (o \rightsquigarrow s)}{\min}  \left( \sum_{h=1}^{H}  \log \left(1+\gamma_h^{-1} \right)\right)
\end{align}
which can also be calculated using the DSP algorithm by setting non-negative edge weights to $\Omega_i^j= \log \left(1+\frac{1}{\gamma_i^j} \right)$ where $\gamma_i^j$ is the SNR between $n_i$ and $n_j$. Once the route is calculated, the $\mathcal{P}^{AF}_{E2E}$ can be calculated by substituting $\bar{R}_{E2E}$ into \eqref{eq:AF_BER}. 

\subsubsection{Maximum E2E Data Rate Routing}
\label{sec:rate}

Our target in this routing scheme is to find a path that maximizes E2E data rate while ensuring a predetermined E2E-BER, $\mathcal{P}^{DF}_{E2E}$, which is non-trivial since the $\mathcal{P}^{DF}_{E2E}$ is coupled with the hop counts as well as $R^{DF}_{E2E}$. Therefore, we first relax the problem by fixing the BER at each edge and then find the optimal route with the maximum $R^{DF}_{E2E}$ as follows
\begin{equation}
\label{eq:DFroute}
o \rightarrow s=\underset{\forall (o \rightsquigarrow s)}{\argmin} \left(\underset{h 
\in H}{\min} \left( R_{h-1}^h \right)\right).
\end{equation}
The problem defined in (\ref{eq:DFroute}) is known as the \textit{widest path} or \textit{bottleneck shortest path} problem and can expeditiously be solved by employing the DSP algorithm via modification of the cost updates \cite{kaibel2006bottleneck}. Depending upon the calculated path, it is necessary to adjust the BER of links to maximize $R^{DF}_{E2E}$ which can be formulated as follows
\begin{equation}
\hspace*{0pt}
 \begin{aligned}
 \label{eq:P1}
 & \hspace*{0pt} \vect{\mathrm{P}_3}:\: \: \underset{\zeta,\vect{\mathcal{P}}}{\max}
& & \hspace*{3 pt} \zeta \\
& \hspace*{0pt} \mbox{$\mathrm{P}_3^1$: }\hspace*{8pt} \text{s.t.}
&&  \zeta  \preceq \vect{R}  \\
   &
  \hspace*{0 pt}\mbox{$\mathrm{P}_3^2$: } & & \mathcal{P}^{DF}_{E2E} \leq \bar{\mathcal{P}}_{E2E}, 
  \end{aligned}
\end{equation} 
which maximizes the minimum rate along the path. In order to have an insight into $\vect{P}_3$, let us consider a two-hop path with channel gains $G_0^1 > G_1^2$. If these links are set to the same BER, $\varepsilon$, we have $R_{E2E}^{DF}=R_1^2$ since $R_0^1 > R_1^2$ due to $G_0^1 > G_1^2$. Thus, $R_{E2E}^{DF}$ is maximized when $R_0^1 = R_1^2$ by manipulating error rates, i.e., $\varepsilon_0^1 < \varepsilon_1^2$. Also notice at the optimal point that the second constraint of $\vect{\mathrm{P}_3}$ is active (i.e., $ \mathcal{P}_{E2E}^{DF} = \bar{\mathcal{P}}_{E2E}$) since enhancing $\mathcal{P}_{E2E}^{DF}$ decrease $R_{E2E}^{DF}$ due to the fundamental tradeoff between rate and error. In light of these, a line search algorithm can find the maximum data rate by minimizing $\vert \mathcal{P}_{E2E}^{DF} - \bar{\mathcal{P}}_{E2E} \vert$ where $\vert x \vert$ denotes the absolute value of $x$. Once the maximum $R_{E2E}^{DF}$ is obtained, BER of each link can be calculated by substituting $R_{E2E}^{DF}$ into \eqref{eq:BER}.


As already explained in the previous section, the optimal path for the maximum E2E data rate is the same with the minimum E2E-BER path that is given by \eqref{eq:AFBERroute}-\eqref{eq:AFBERroute3}. Once the route is calculated by the DSP algorithm, $R_{E2E}^{AF} $ can be calculated by substituting $\bar{\mathcal{P}}_{E2E}$ into \eqref{eq:AFrate}.  

\subsubsection{Minimum Power Consumption}
\label{sec:pwr}
The UOWN nodes have limited energy budget that has a significant impact on the network lifetime. Accounting for the monetary cost and engineering hardship of battery replacement, an energy efficient routing technique is essential to minimize the energy consumption of multihop communications. For a given pair of E2E data rate, $\bar{R}_{E2E}$, and BER constraint, $\bar{\mathcal{P}}_{E2E}$, the objective of this routing scheme is to find the optimal path which provides the minimum total transmission power. Because of the inextricably interwoven relationship between hop counts, $R_{E2E}$ and $\mathcal{P}_{E2E}$ finding the most energy efficient route falls within the class of mixed-integer non-linear programming (MINLP) problems which is known to be NP-Hard. Therefore, we propose fast yet efficient suboptimal solutions for both DF and AF relaying. For the DF relaying, E2E-Rate constraint can be satisfied by exactly setting the data rate of each edge weight to $\bar{R}_i^j=\bar{R}_{E2E}, \forall i,j$. Then, edge weights are calculated using \eqref{eq:DF_Pt} by fixing the BER at each hop. In this way, the minimum power cost path can be obtained by using the DSP algorithm. Once the route is calculated, optimal BER that minimizes the total energy consumption can obtained by solving $\vect{\mathrm{P}_1}$.

Compared to the DF relaying, finding the optimal route for the AF relaying is more complicated. As can be seen from the constraint of \eqref{eq:P2}, the optimal transmission power varies with the number of hops and channel gains along the routing path, which prevents using the DSP algorithm to find the optimal path. Alternatively, we consider selecting the path with minimum $P_t^*$ among a number of feasible path that satisfies constraints which can be verified by \eqref{eq:AF_BER} and \eqref{eq:AFrate}. In this regard, $k-$shortest path (KSP) algorithm can be employed as a generalization of the DSP such that it also finds $k$-paths with the highest $\gamma_H$s in the order of $\mathcal{O}\left( \vert\mathcal{E}\vert + \vert\mathcal{V}\vert \log\left(\vert\mathcal{V}\vert\right) +k \right)$ \cite{KSP}. For feasible paths that satisfy $ R_{E2E}^{AF} \geq \bar{R}_{E2E}^{AF}$ and $\mathcal{P}^{AF}_{E2E} \leq \bar{\mathcal{P}}_{E2E}$, we solve the problem $P_2$ in \eqref{eq:P2} and select the path with minimum $H P_t^*$ as the final route.  

\subsection{\textbf{Li}ght \textbf{Pa}th \textbf{R}outing (\textbf{LiPaR}): A Distributed Routing Scheme}
Since node location information is a prerequisite for pointing to establish links, geographic routing schemes can be regarded naturally potential methods for UOWNs. Even though there exists many successful geographic routing protocols designed for omni-directional RF communications in terrestrial wireless sensor networks, they cannot be directly applied to UOWNs because of the directed nature of optical wireless communications and hostile underwater environment. Accordingly, we propose a distributed routing algorithm, namely \textbf{Li}ght \textbf{Pa}th \textbf{R}outing (LiPaR), that only requires the location information of the sink and neighboring nodes. LiPaR does not rely upon a PAT mechanism and assumes that each node directs its pointing vector to the closest sink. Each forwarder node first determines the nodes within its communication range, which is certainly subject to the fundamental range-beamwidth tradeoff of UOWCs. Therefore, the set of feasible relays within the neighborhood of $n_s$ can be given by
\begin{equation}
\label{eq:neighbor}
\aleph_i=\{j \vert\: D_i^j \geq \vert\vert \vect{\ell_j} -\vect{\ell_i} \vert \vert, \: \theta_{min} \leq \varphi_i^j \leq \theta_{max}, \forall j \}, \forall i
\end{equation}
which is defined based on a predetermined data rate and BER pair at each hop. In what follows, the forwarder for the next hop is chosen among $n_j \in \aleph_s, \forall j$ as follows 
\begin{equation}
\label{eq:forwarder}
h_{i+1}=\underset{j}{\argmax} \left\{(1-\mathcal{P}_s^j) \times \vert\vert \vect{\ell_j} -\vect{\ell_i} \vert \vert, \forall j \in \aleph_i \right\}
\end{equation}
which maximizes the distance progress by also taking the link quality into account. Notice that $\mathcal{P}_s^j$ varies both with distance and angle between $n_j$ and pointing vector of $n_s$, $\varphi_s^j$. Let us consider two candidate nodes $n_1$ and $n_2$ with equal Euclidean distance but different angles $\varphi_s^1> \varphi_s^2$. In case of $n_2$ is chosen as a next forwarder, the transmitter needs to adjust its divergence angle to a tighter beamwidth, which maximizes the average distance progress since $\mathcal{P}_s^j$ is minimized. Following \eqref{eq:forwarder}, the current forwarder adapts its beamwidth to cover the next one based on \eqref{eq:theta_no_pat} and transmit packets after informing the next node about its decision. This procedure is repeated along the path until one of the sink node is reached.

\begin{figure}[b!]
\centering
\begin{subfigure}[b]{0.32\columnwidth}
\includegraphics[width=\columnwidth]{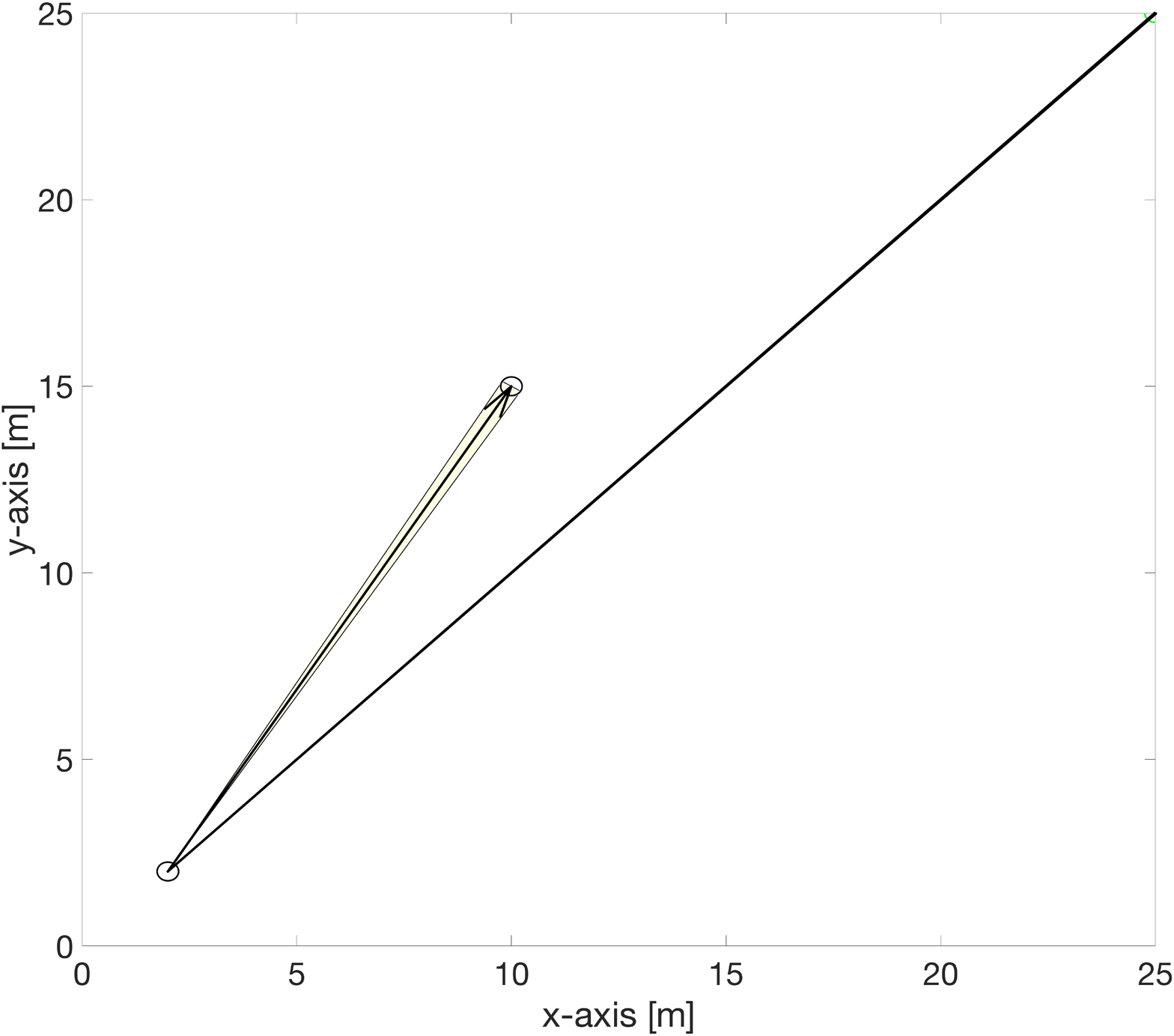}
\caption{}
\label{fig:a}
\end{subfigure}
\begin{subfigure}[b]{0.32\columnwidth}
\includegraphics[width=\columnwidth]{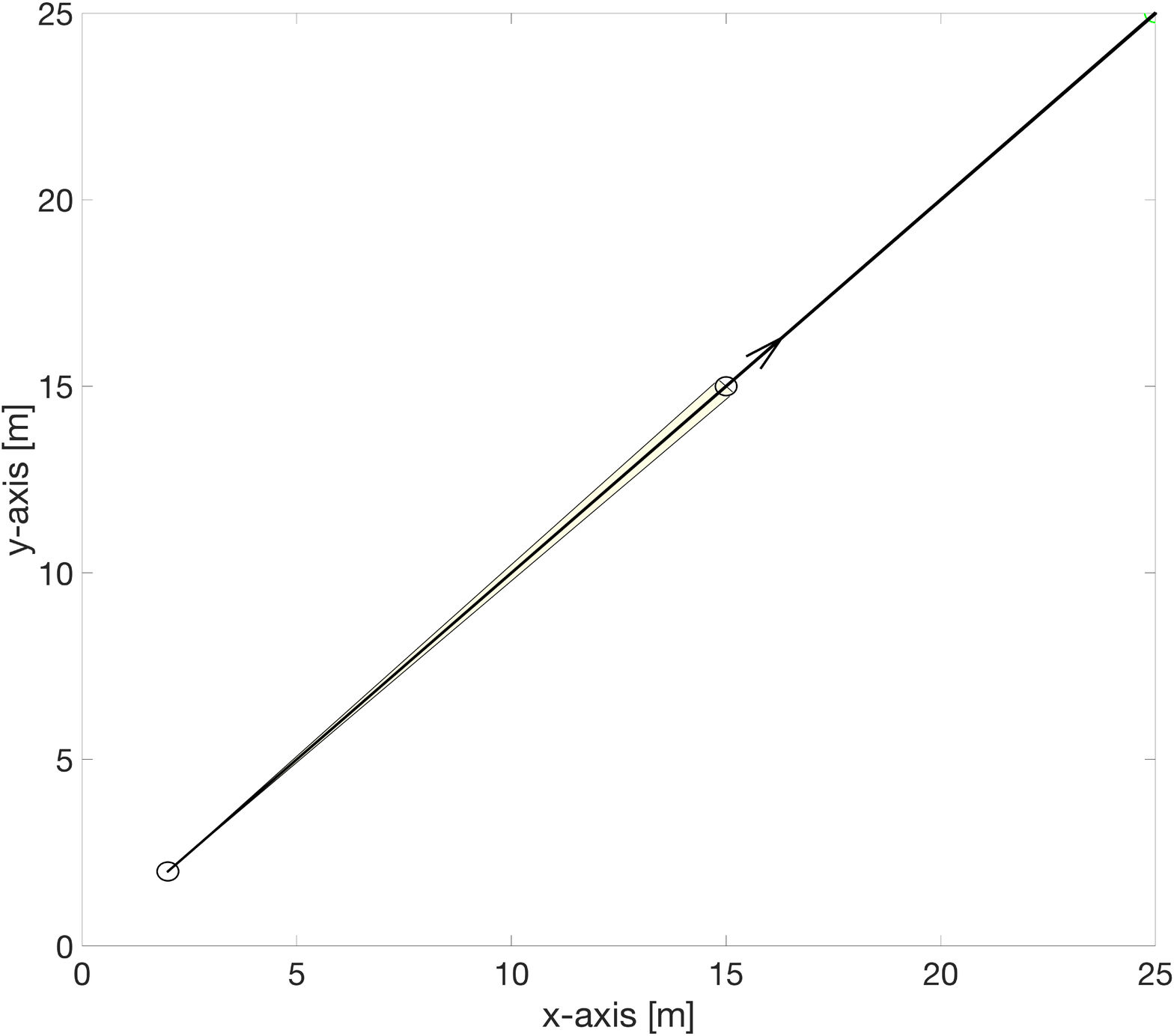}
  \caption{}
  \label{fig:b}
\end{subfigure}
\begin{subfigure}[b]{0.32\columnwidth}
\includegraphics[width=\columnwidth]{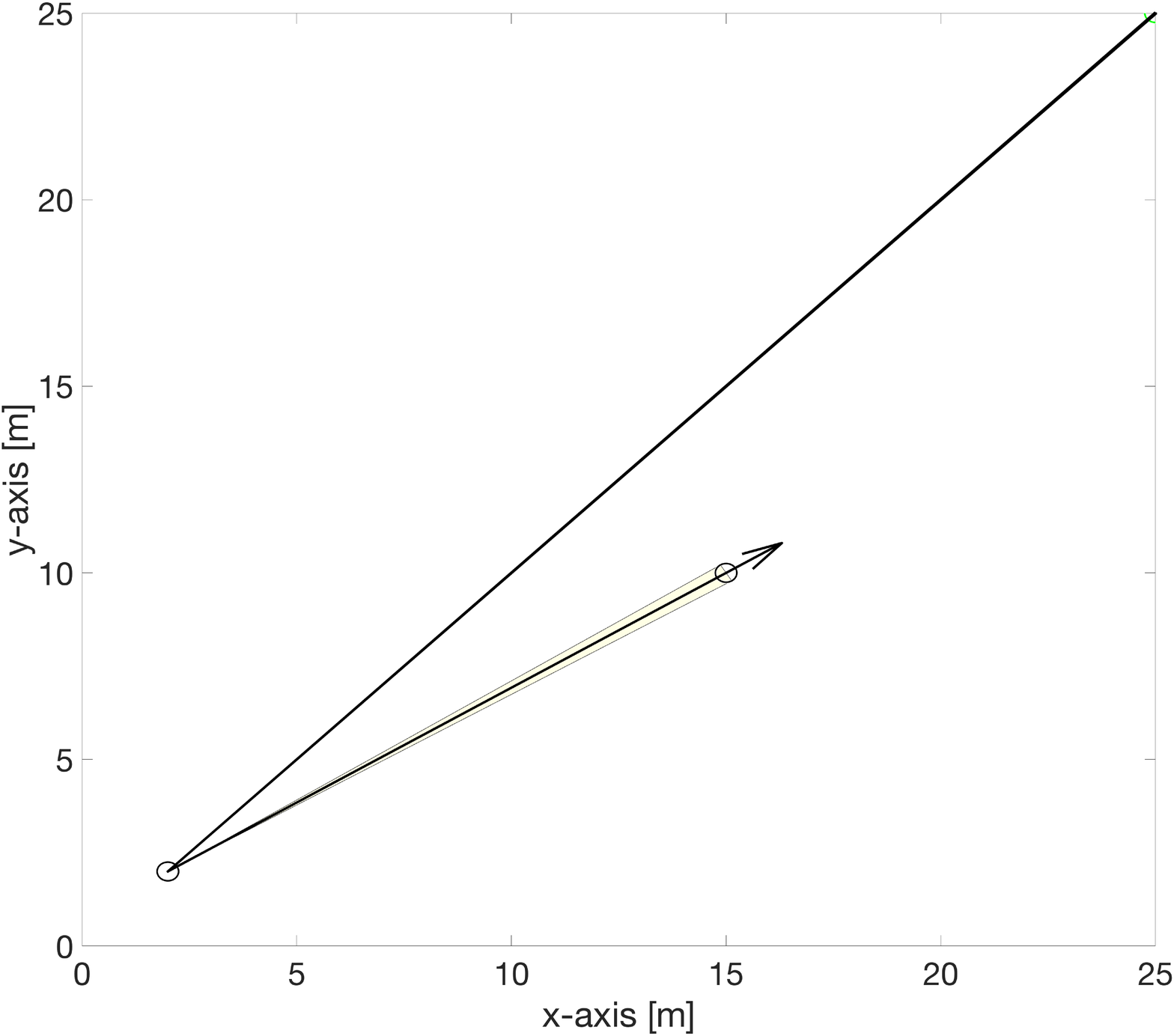}
  \caption{}
  \label{fig:c}
\end{subfigure}

\centering
\begin{subfigure}[b]{0.32\columnwidth}
\includegraphics[width=\columnwidth]{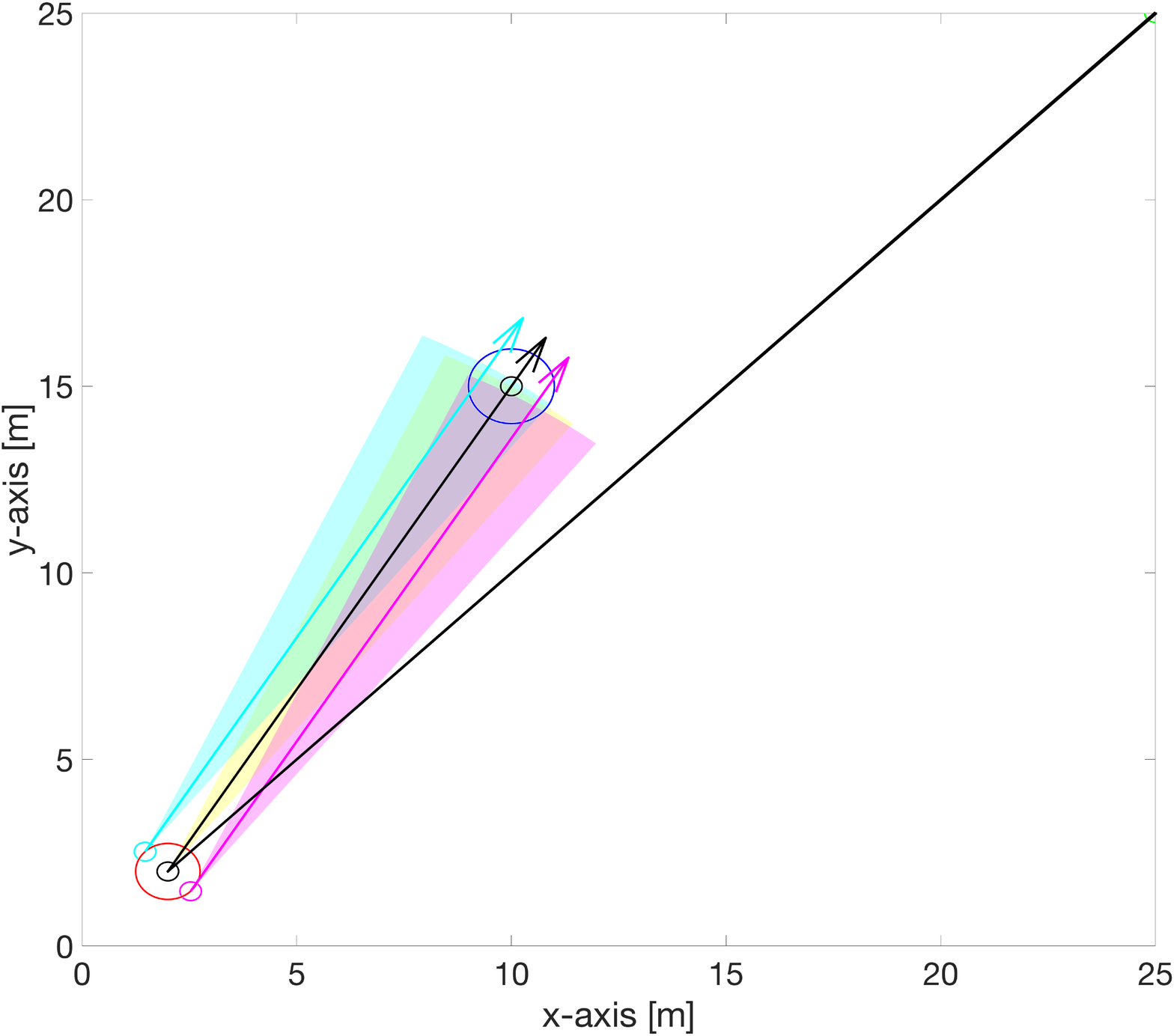}
\caption{}
\label{fig:d}
\end{subfigure}
\begin{subfigure}[b]{0.32\columnwidth}
\includegraphics[width=\columnwidth]{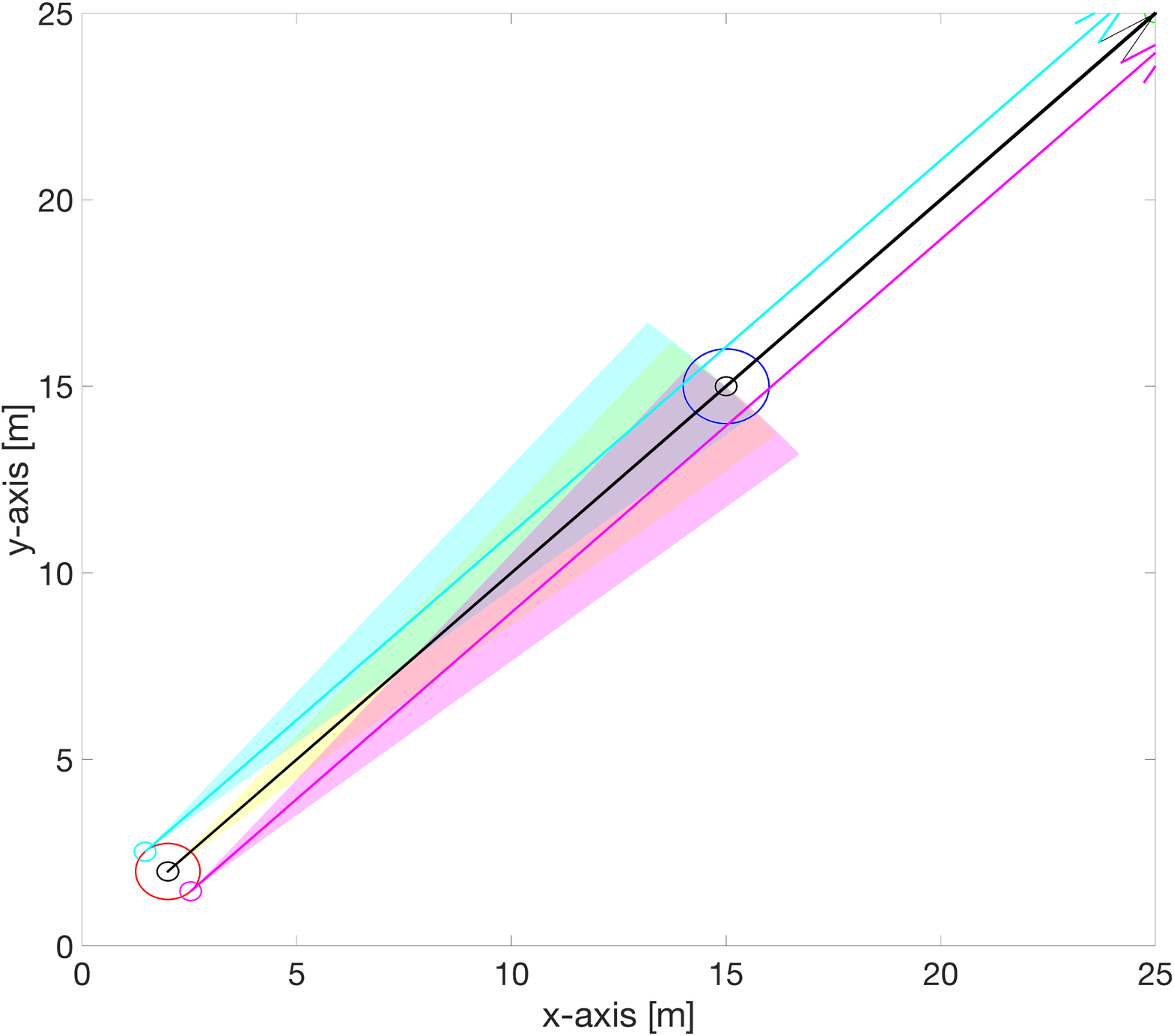}
  \caption{}
  \label{fig:e}
\end{subfigure}
\begin{subfigure}[b]{0.32\columnwidth}
\includegraphics[width=\columnwidth]{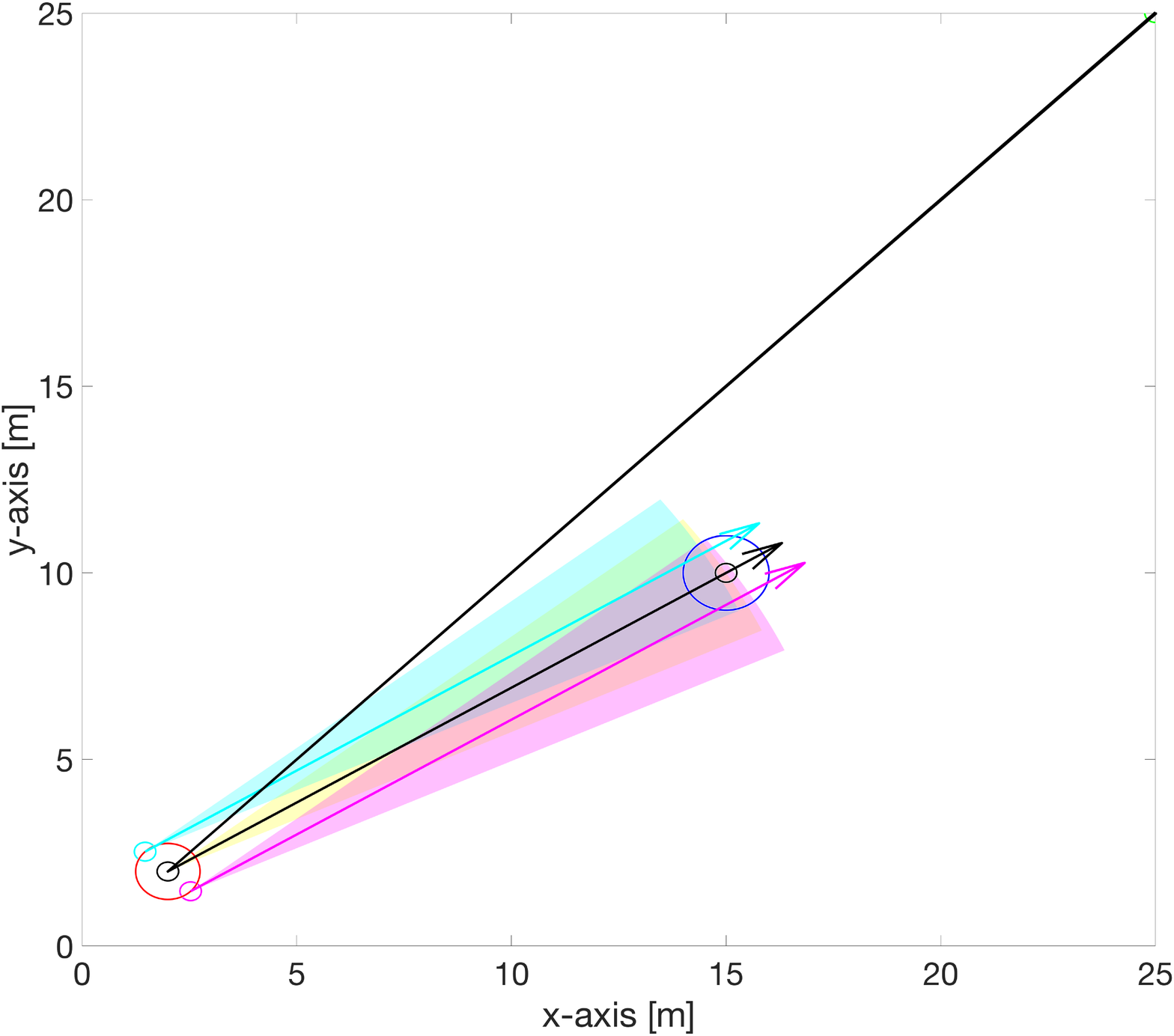}
  \caption{}
  \label{fig:f}
\end{subfigure}

\centering
\begin{subfigure}[b]{0.32\columnwidth}
\includegraphics[width=\columnwidth]{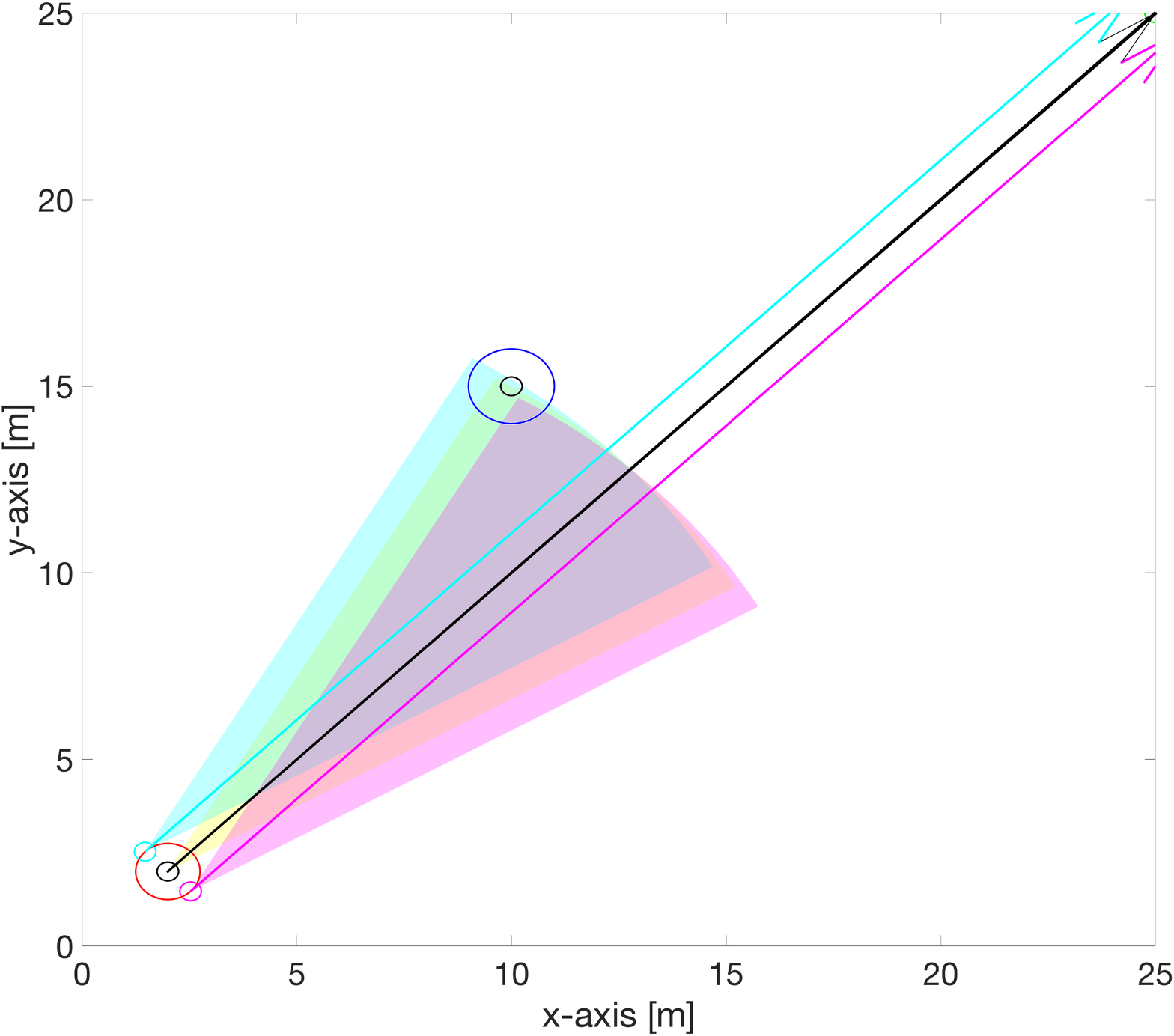}
\caption{}
\label{fig:g}
\end{subfigure}
\begin{subfigure}[b]{0.32\columnwidth}
\includegraphics[width=\columnwidth]{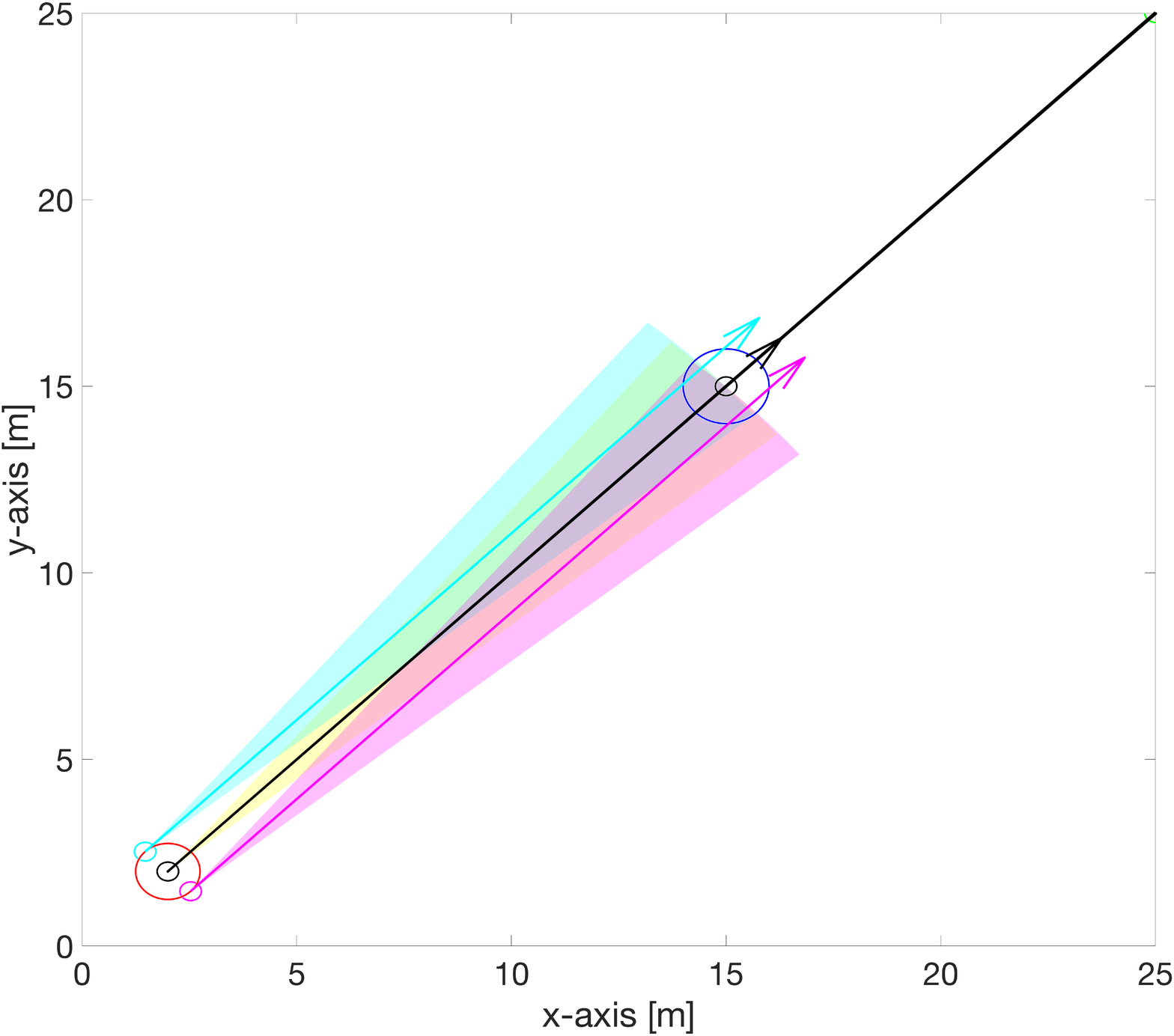}
  \caption{}
  \label{fig:h}
\end{subfigure}
\begin{subfigure}[b]{0.32\columnwidth}
\includegraphics[width=\columnwidth]{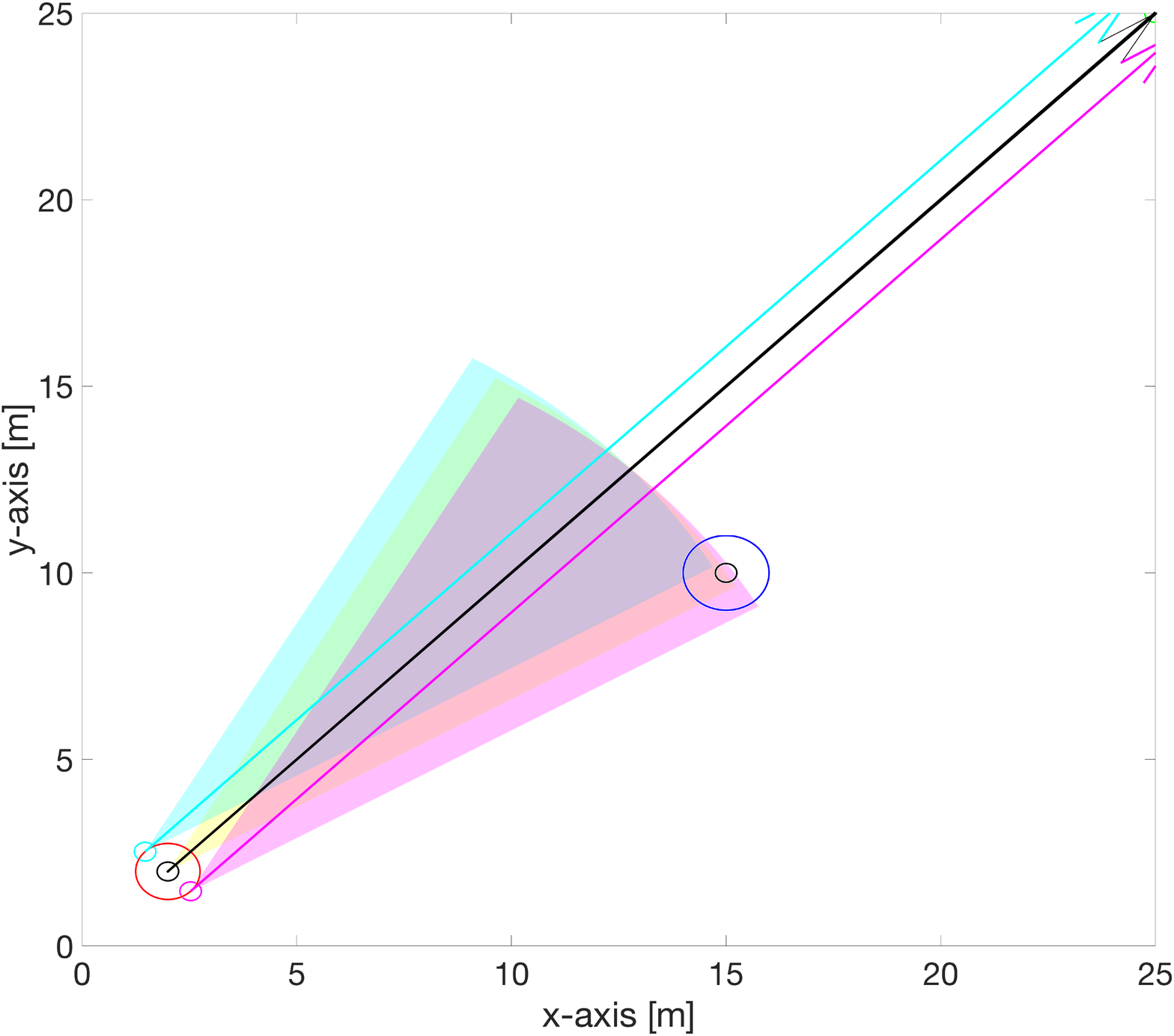}
  \caption{}
  \label{fig:i}
\end{subfigure}
  \caption{Beamwdith calculations for three pointing schemes: 1) PAT mechanism with perfect locations (a-c), 2) PAT mechanism under location uncertainty (d-f), and 3) Only adaptive beamwidth without a PAT mechanism (g-i).}
  \label{fig:pat}
\end{figure}

\begin{figure*}[t]
\centering
\begin{subfigure}[b]{0.32\textwidth}
\includegraphics[width=\columnwidth]{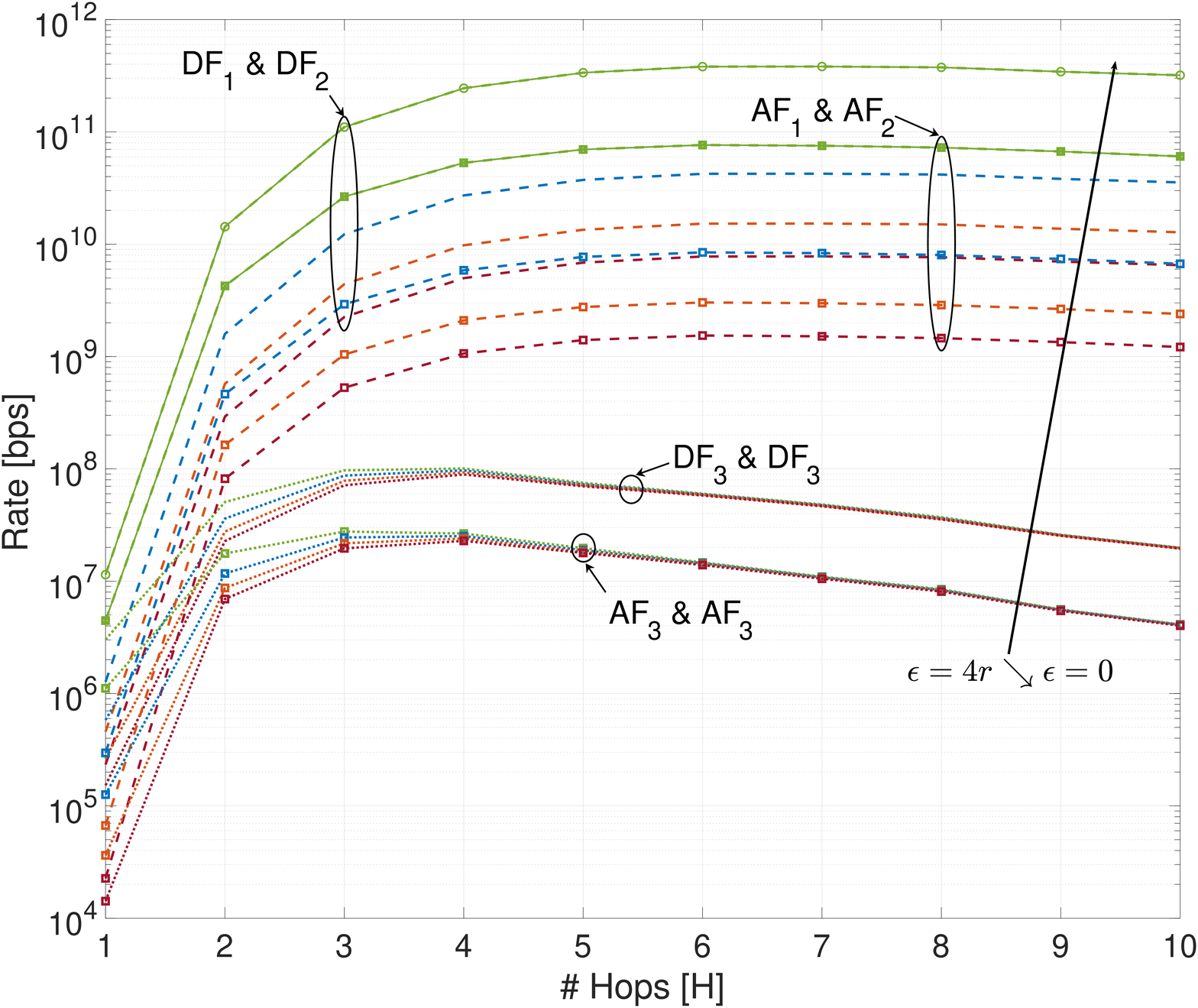}
\caption{$R_{E2E}$ vs. hop counts}
\label{fig:eps_anl_rate}
\end{subfigure}
\begin{subfigure}[b]{0.32\textwidth}
\includegraphics[width=\columnwidth]{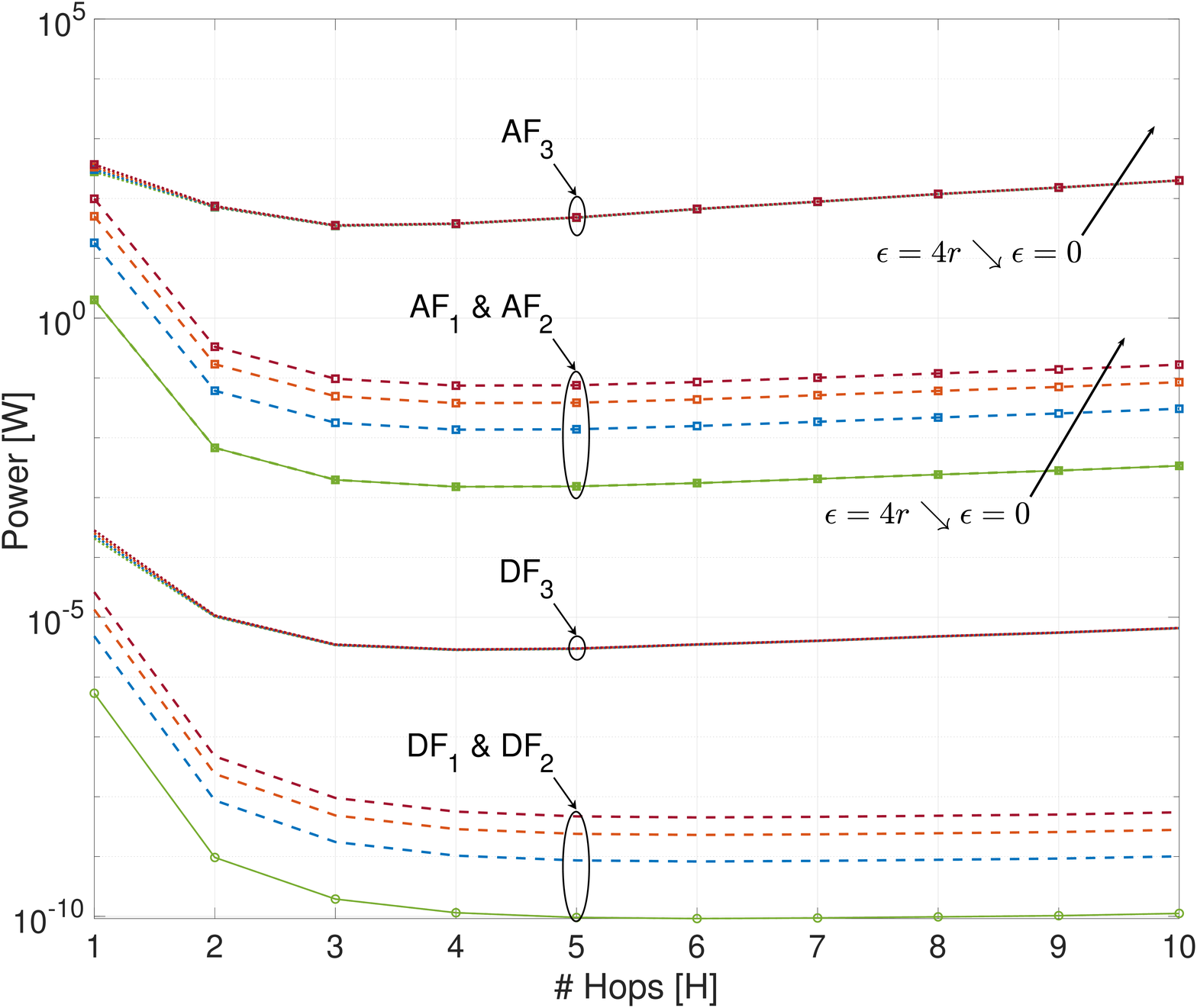}
\caption{Total transmit power vs. hop counts}
  \label{fig:eps_anl_pwr}
\end{subfigure}
\begin{subfigure}[b]{0.32\textwidth}
\includegraphics[width=\columnwidth]{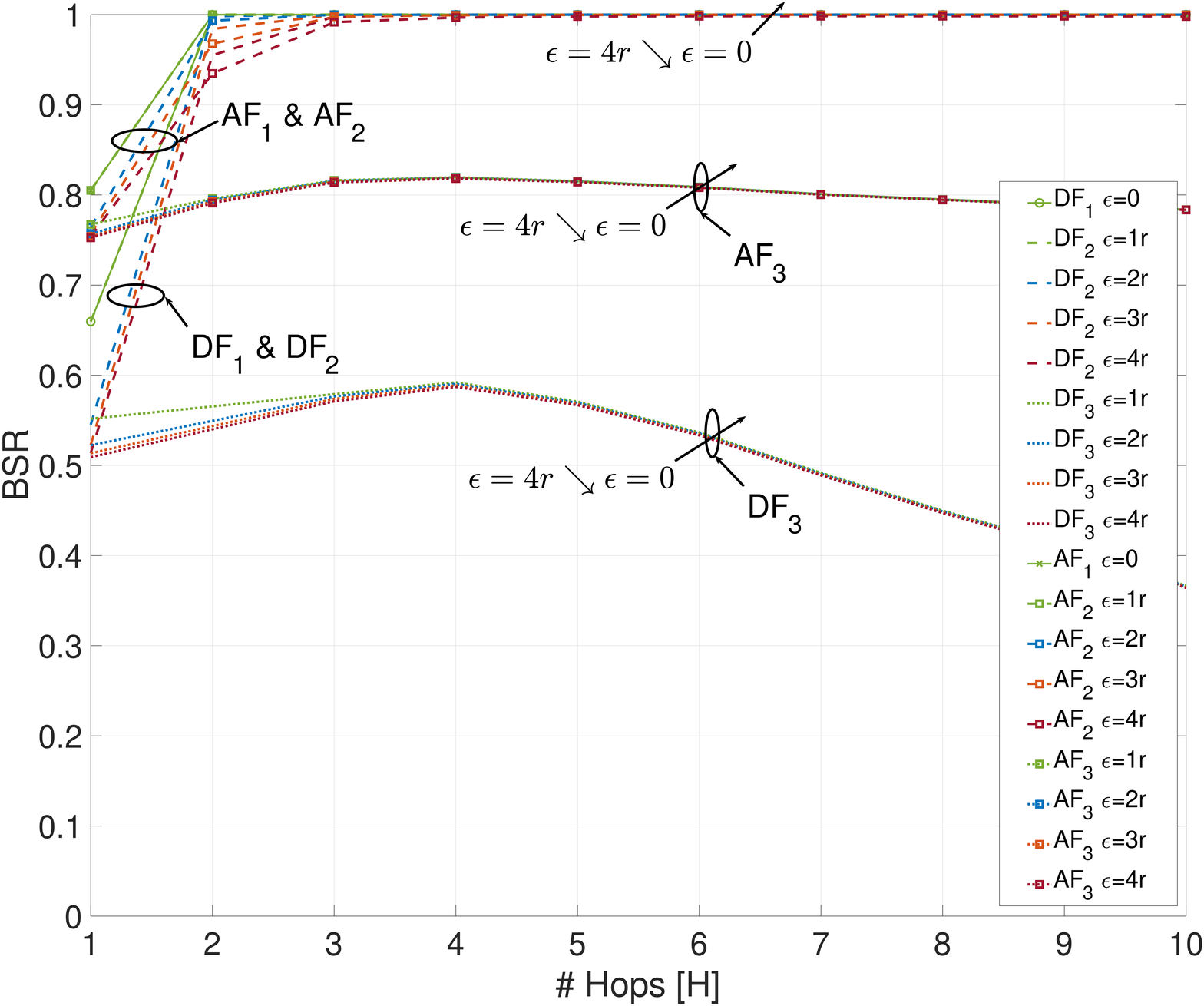}
\caption{BSR vs. hop counts}
  \label{fig:eps_anl_bsr}
\end{subfigure}
\caption{Impacts of location uncertainty on rate, power, and BSR under different pointing cases.}
  \label{fig:eps_anl}
\end{figure*}

\begin{figure*}[t]
\centering
\begin{subfigure}[b]{0.32\textwidth}
\includegraphics[width=\columnwidth]{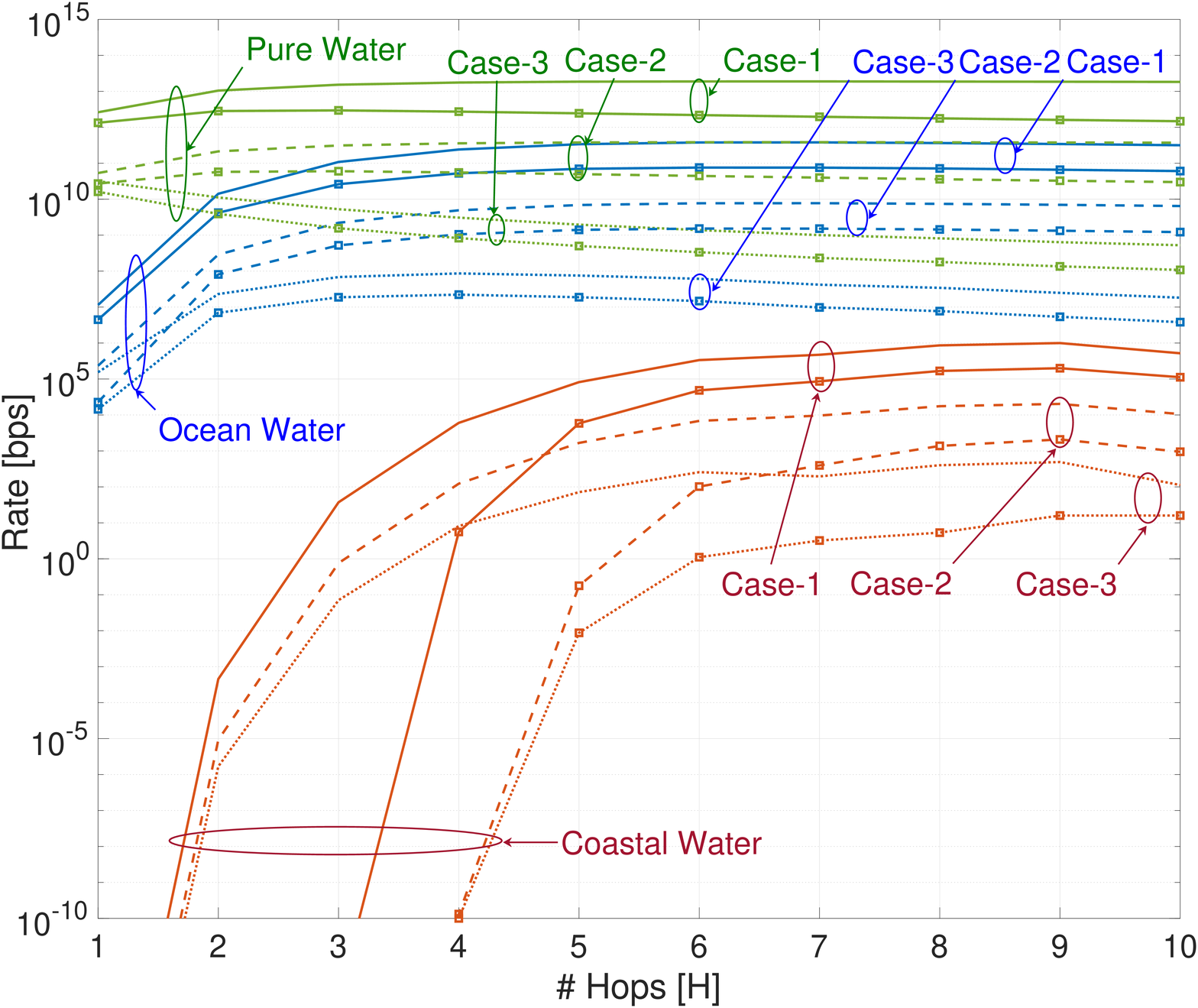}
\caption{$R_{E2E}$ vs. hop counts}
\label{fig:wtr_anl_rate}
\end{subfigure}
\begin{subfigure}[b]{0.32\textwidth}
\includegraphics[width=\columnwidth]{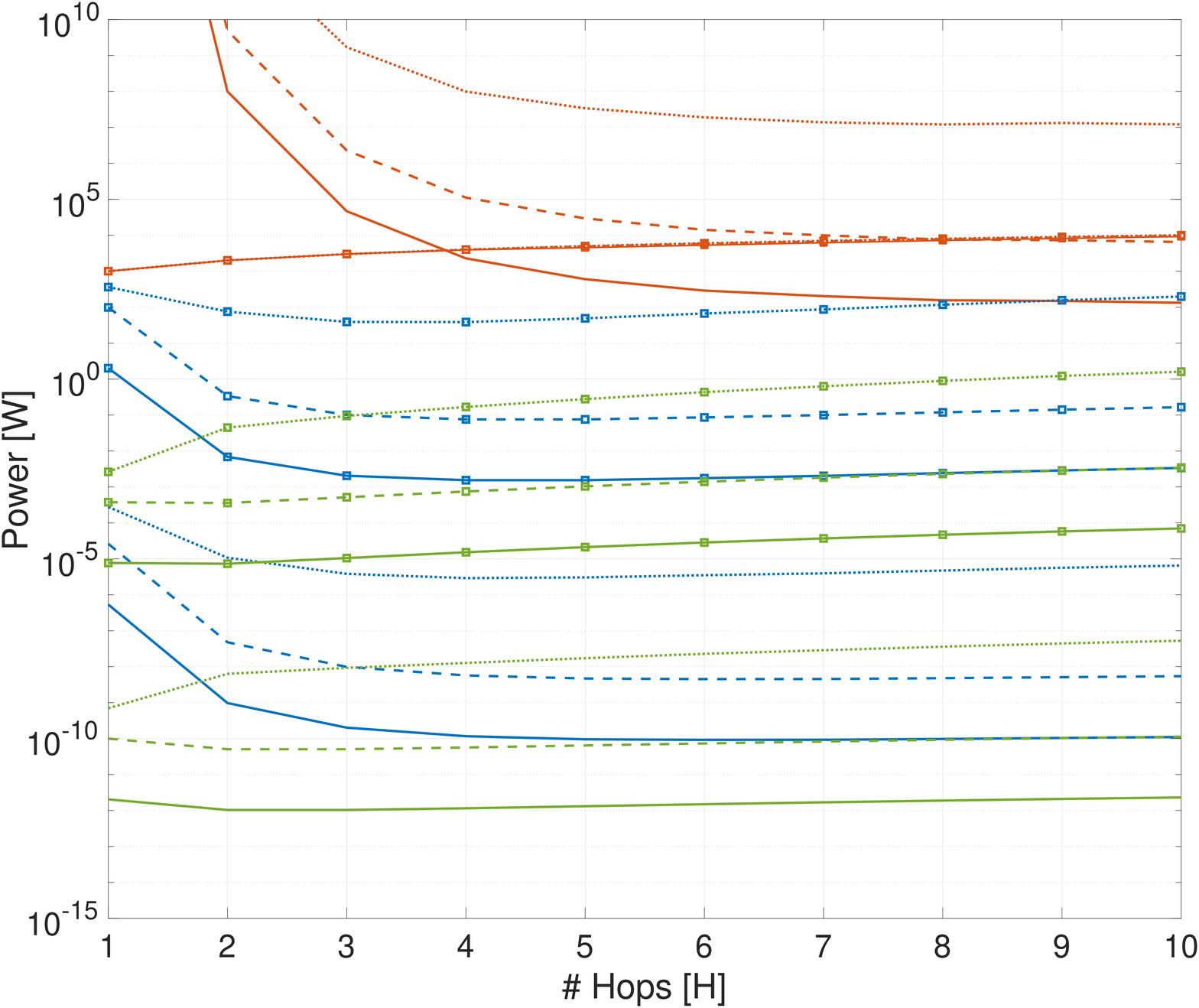}
\caption{Total transmit power vs. hop counts}
  \label{fig:wtr_anl_pwr}
\end{subfigure}
\begin{subfigure}[b]{0.32\textwidth}
\includegraphics[width=\columnwidth]{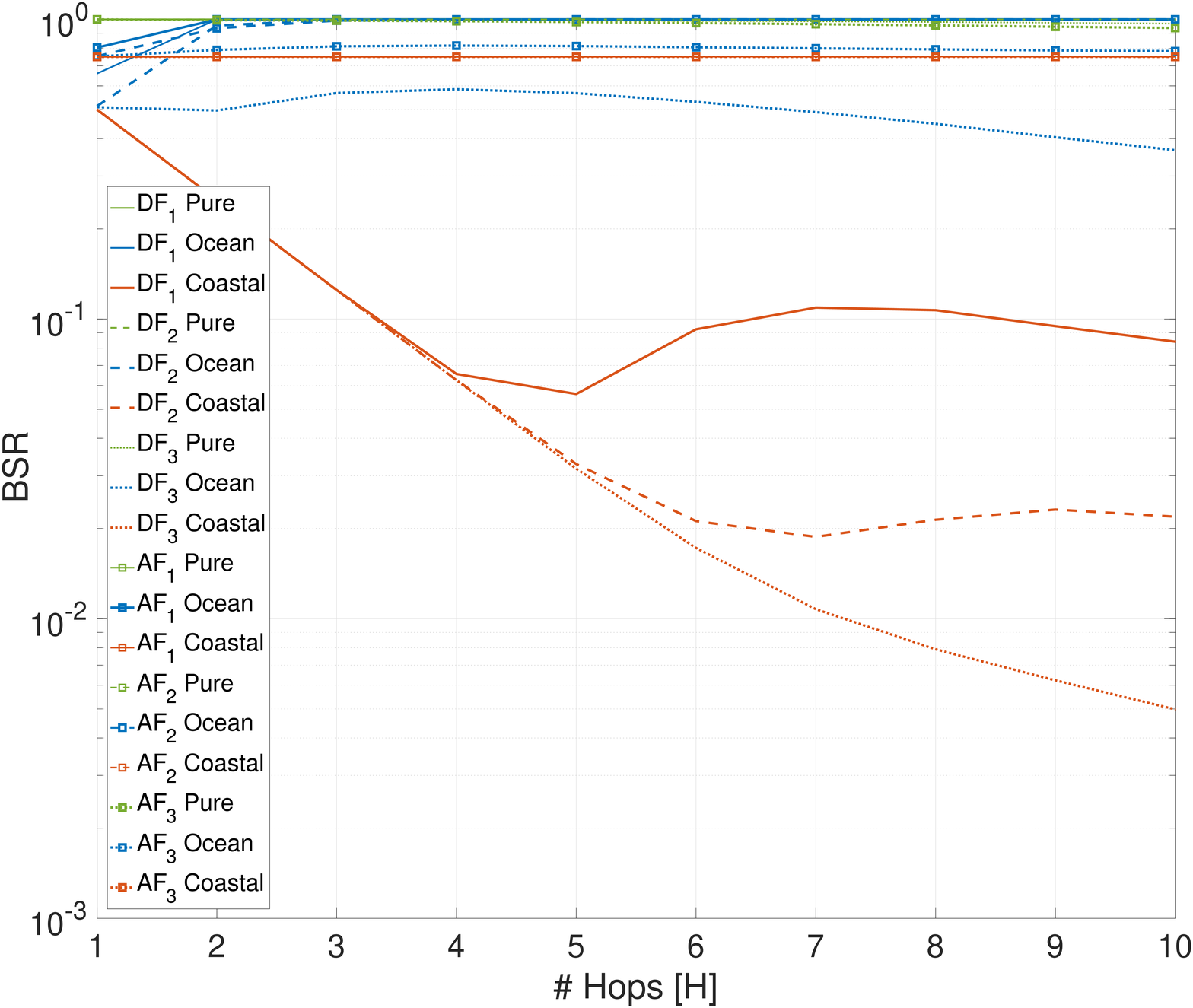}
\caption{BSR vs. hop counts}
  \label{fig:wtr_anl_bsr}
\end{subfigure}
\caption{Impacts of water types on rate, power, and BSR under different pointing cases.}
  \label{fig:wtr_anl}
\end{figure*}

\begin{table}[t]
\centering
\caption{Table of Parameters}
\label{tab:par}
\scriptsize 
\resizebox{\columnwidth}{!}{%
\begin{tabular}{|l|l|l|l|l|l|}
\hline
Par.         & Value       & Par.          & Value           & Par.          & Value         \\ \hline
$P_t$    & $10$ $\rm{mW}$ &$\hslash$ & $6.62\mathrm{E}-34$   & $\bar{\mathcal{P}}_{E2E}$   & $1\mathrm{E}{-5}$   \\ \hline
$\eta_x$ & $0.9$     &$c$& $2.55\mathrm{E}{8}$ $\rm{m/s}$&$\bar{R}_{E2E}$      & $1$ $\rm{Gbps}$        \\ \hline
$ \eta_r$& $0.9$     &$\lambda$ & $532\mathrm{E}-9$     & $M$  & $3$    \\ \hline
$\eta_d$ & $0.16$    &$e(\lambda)$ & $0.1514$           & $N$ & $60$     \\ \hline
$A$      & $5$ $\rm{cm}$  &$f_h(P_{dc})$ & $1\mathrm{E}{6}$       & $f_h(P_{bg})$ & $1\mathrm{E}{6}$      \\ \hline
$ T$   & $1$ $\rm{ns}$  &$\theta_{\min}$ & $10$ mrad       & $\theta_{\max}$ & $0.25$ rad      \\ \hline
$r$   & $0.25$ $\rm{m}$  & $\epsilon$ &  $0.75$ $\rm{m}$       & $P_{n}$ & $-84$ dBm      \\ \hline
\end{tabular}%
}
\end{table}

\section{Numerical Results}
\label{sec:res}
In this section, we provide the performance evaluations using the default parameters listed in Table \ref{tab:par} which is mainly drawn from \cite{Arnon10underwater}. Simulations are conducted on Matlab and presented results are averaged over $10,000$ random realizations.

\subsection{Validation of Adaptive Beamwidth Calculations}
 Before presenting the performance evaluations, let us illustrate the validation of the beamwidth calculations provided in Section \ref{sec:PAT}. In Fig. \ref{fig:pat}, we demonstrate the adaptive beamwidths between a transmitter $i$ located at $\vect{\ell}_i=[2 \: 2]$ and a randomly located receiver $j$ for three different cases: Case-1) PAT with perfect location estimates [c.f. Fig. \ref{fig:a}-Fig. \ref{fig:c}], Case-2) PAT under location uncertainty [c.f. Fig. \ref{fig:d}-Fig. \ref{fig:f}], and Case-3) Adaptive beamwidths in the absence of PAT  [c.f. Fig. \ref{fig:g}-Fig. \ref{fig:i}]. Thanks to the availability of the actual node locations, Case-1 is able to tune the divergence angle for a minimum beamwidth to barely cover the receiver node, which naturally delivers the best performance. Compared to the first case, Case-2 requires larger beamwidths to compensate the location uncertainty of the transceivers, which degrade the achievable performance. In the last case, the transmitter is directed toward to the sink node located at $\vect{\ell}_s=[25 \: 25]$ and required to adjust its beamwidth to cover a receiver located far way from the transmitter-receiver trajectory. Accordingly, performance degrades as the beamwidth increases to establish a link towards a receiver with a higher $\varphi_i^j$. In the remainder of this section, these three cases will be indicated by means of superscripts, e.g., DF$_i$ refers to the DF relaying in Case-i, $i=1,2,3$.   

\subsection{E2E Performance Evaluation of DF and AF Relaying}
For the E2E performance evaluation of DF and AF relaying schemes, we consider a routing path with $H$ hops to reach a sink node $200 m$ far away from the source node. Although the length of hops are the same, the angle between the relays and source-sink trajectory is random. Fig. \ref{fig:eps_anl} demonstrate the impacts of location uncertainty on rate, power, and BSR under different pointing schemes. In Fig. \ref{fig:eps_anl}, previous discussions are validated as Case-$i$ always delivers a superior performance compared to Case-$j$, $j>i$. As the location uncertainty increases, the performance degrades for all cases. For $H<4$, increasing the hop count serve as a remedy to compensate the distance related losses. However, for $H\geq 4$, decreasing the hop lengths by increasing the hop count triggers the distance-beamwidth tradeoff. That is, divergence angle becomes significant even for a closer node because perpendicular distance and the distance between receiver and pointing vector are close. Therefore,  Fig. \ref{fig:eps_anl} exhibits a performance enhancement until $H=4$ and degradation after that. This behavior has also an impact on the location uncertainty such that Case-3 is not effected from location uncertainty driven beamwidth enlargement because aforementioned effect makes this additional divergence angle increase less significant.

\begin{figure}[t!]
\includegraphics[width=\columnwidth]{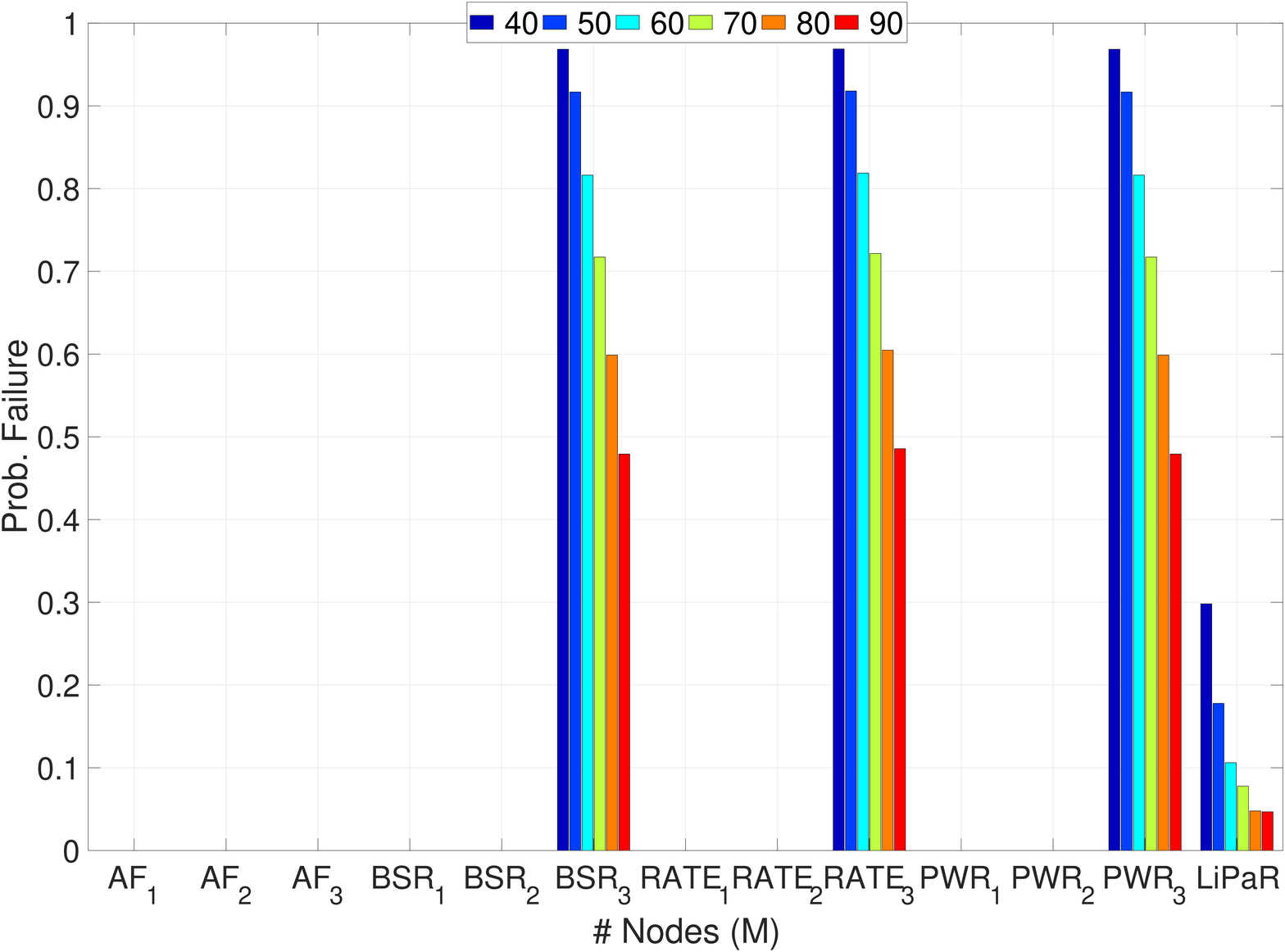}
\caption{Probability of failures under different network densities.}
\label{fig:fail}
\end{figure}
\begin{figure}[t!]
\includegraphics[width=\columnwidth]{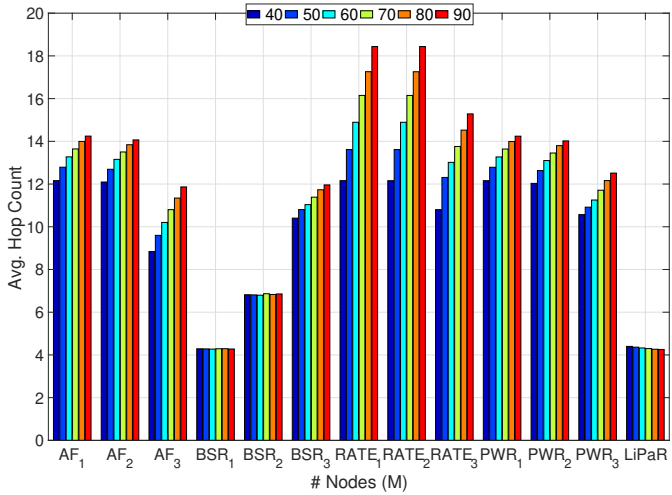}
\caption{Average hop counts under different network densities.}
\label{fig:hop_counts}
\end{figure}

The E2E-Rate performance is shown in Fig. \ref{fig:eps_anl_rate} where DF relaying has better rates AF relaying thanks to its ability to detecting and regenerating the bits at each hop. Unlike the first two cases, rates of Case-3 monotonically reduces for $H\geq 4$ due to the low rates as a result of larger beamwidths as explained above. In  Fig. \ref{fig:eps_anl_pwr}, the DF relaying is also shown to be more energy efficient than the AF relaying, which is expected since the noise propagation along the path necessitates a higher transmission power. We should note that schemes requiring transmission powers more than $1$ $W$ may not be feasible because of the power budget limitations of underwater sensor nodes. Reminding that we merely count for the transmission power, the AF relaying may be far energy efficient than the DF relaying because it does not need energy and computational power hungry conversion and modulation circuitry. Finally, Fig. \ref{fig:eps_anl_bsr} shows that the AF relaying provides a better BSR performance than the DF relaying, this is indeed caused by main difference between DF and AF relaying. Unlike Case-3, existence of a PAT mechanism eliminates this difference and provide a desirable performance starting from $H=3$.
\begin{figure}[t!]
\includegraphics[width=\columnwidth]{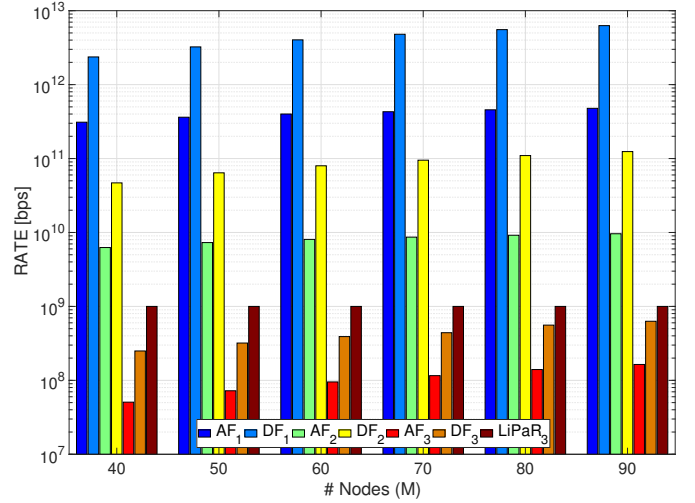}
\caption{E2E-Rates under different network densities.}
\label{fig:rate}
\end{figure}
\begin{figure}[t!]
\includegraphics[width=\columnwidth]{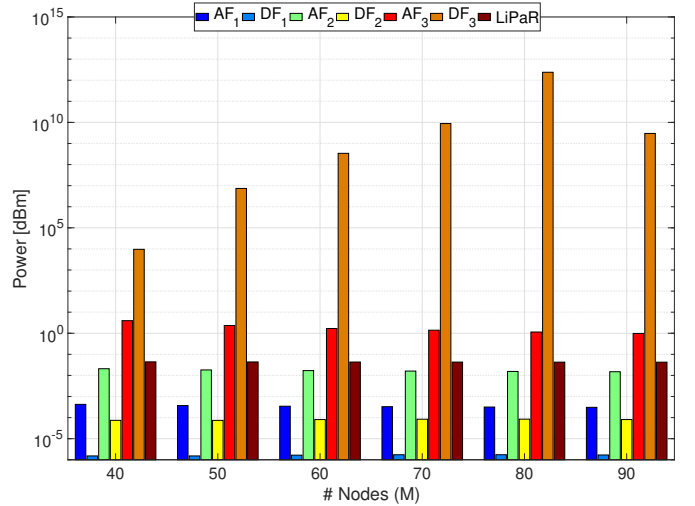}
\caption{Total power consumptions under different network densities.}
\label{fig:pwr}
\end{figure}

In Fig. \ref{fig:wtr_anl}, we show how different cases of rate, power, and BSR is affected from various water types; pure water ($e(\lambda)=0.056$), ocean water ($e(\lambda)=0.151$), and coastal water ($e(\lambda)= 0.398$). Apparently, the rate, power, and BSR performance degrades as the density of the water particulates increases. We also still observe that Case-$i$ always delivers a superior performance compared to Case-$j$, $j>i$. In Fig. \ref{fig:wtr_anl_rate}, we observe that the DF relaying rate surpasses that of the AF relaying in all cases and water types. Interestingly, the benefits of multihop communications become significant as the hostility of the underwater environment increases. Excluding the coastal water,  Fig. \ref{fig:wtr_anl_pwr} shows that the DF relaying  is more energy efficient than the AF relaying. In Fig. \ref{fig:wtr_anl_bsr}, we observe that coastal water deteriorates the BSR performance in a great extent. 

Overall,  Fig. \ref{fig:eps_anl} and Fig. \ref{fig:wtr_anl} clearly shows that UOWC urges multihop communication which is established on PAT mechanism along with accurate node locations. Otherwise, E2E performance can be affected from harsh channel conditions and fundamental tradeoff between range and beamwidth. This also necessitates efficient routing protocols which are examined next.

\begin{figure}[t!]
\includegraphics[width=\columnwidth]{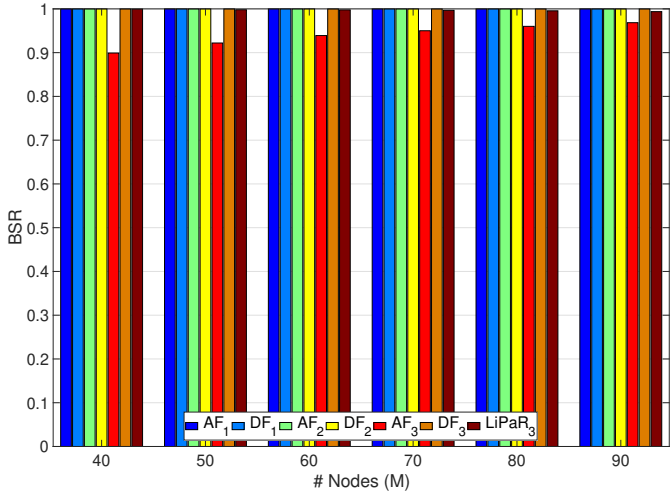}
\caption{E2E-BSRs under different network densities.}
\label{fig:bsr}
\end{figure}

\subsection{Comparison of the Proposed Protocols}
This section compares the performance of the proposed centralized and distributed routing schemes. We consider a network area of $100 m \times 100 m$ where nodes are uniformly distributed. At each realization, the source node is located at a random location at the sea bed. On the other hand, sink locations are arranged to be centered at the sea surface with equidistant intervals. 

Fig. \ref{fig:fail} shows the percentage of failure in finding a route between source and one of the sink nodes. In the $x$-axis, we denote the routing schemes with their objective and underlying pointing method, e.g., RATE$_i$ corresponds to maximum rate routing based on Case-$i$ pointing method, $i=1,2,3$. It is obvious from Fig. \ref{fig:fail} that Case-1 and Case-2 was always able to find a path thanks to their PAT mechanism. However, Case-3 suffers from connectivity of the network which is degraded in a great extent due to the range-beamwidth tradeoff. Nonetheless, increasing the node density helped to reduce failures as a result of increased connectivity. Finally, LiPaR is shown to deliver a good performance in terms of finding a path towards one of the sinks. Reminding that LiPaR cannot guarantee an E2E performance, having Case-3 with higher failures is because of the infeasibility to assure an E2E performance.

Fig. \ref{fig:hop_counts} shows the average hop count of the calculated paths. A common behavior for all centralized scheme is that number of hops increase with node density. This is expected since having shorter hop lengths provides a better performance in all terms of the performance. However, we observe a slight decrease in average hop count of LiPaR due to its hop-by-hop nature. As the node density increases, it is more probable to reach the sink nodes at lower number of hops. 

Fig. \ref{fig:rate} demonstrates the E2E-Rate performance of different routing schemes described in Section \ref{sec:rate}. Results show that the DF relaying provides a higher E2E-Rate in all cases. The rate is also monotonically increases with the node density due to better connectivity and higher performance at individual links. Again, LiPaR delivers a better performance than Case-3 and it is always fixed to 1 GGbps because candidate nodes are defined based on this fixed rate.

Fig. \ref{fig:pwr} depicts the E2E total power consumption of different routing schemes described in Section \ref{sec:rate}. Results show that the DF relaying requires less total power than the AF relaying. However, the DF relaying can still be more power hungry because of the OEO conversion and signal processing. The one exception to this relation is the Case-3 where AF relaying demands less transmit power. Nonetheless, Case-3 requires infeasibly high transmit power, which makes LiPaR a better solution as it requires less power and operates in a distributed manner.

The BSR performance of different routing schemes described in Section \ref{sec:bsr} is demonstrated in Fig. \ref{fig:bsr} where all schemes reaches desirable levels except the AF$_3$ which suffers from the absence of a PAT mechanism and it is not possible to compensate the loss by detecting and regenerating the signals as in DF$_3$.

Finally, we have evaluated the performance metrics for various number of sinks, i.e., $N \in [1,5]$. Since the centralized schemes already have the entire network information, we did not observe a significant change in their performance for different $N$. However, LiPaR has shown a considerable performance enhancement in terms of failures. The percentage of failures for $N \in [1,5]$ is recorded as $[0.25 \: 0.18 \: 0.10 \: 0.08 \: 0.05]$ with a slight change in average number hops. Indeed, high number of nodes make it possible to provide a connection opportunity to relays near the surface such that node clusters who can reach these nodes is granted access to the sinks.

\section{Conclusions}
\label{sec:conc}
Multihop communication is a promising solution to alleviate the short communication range limitation of UOWCs. In particular, a proper operation of UOWNs relies upon the degree of connectivity which can be improved via multihop communications. In this regard, analyzing the E2E performance of multihop UOWC  is necessary to gain insight into the UOWNs. In this paper, we accordingly investigated multihop UOWCs underlying two prominent relaying techniques: the DF and AF relaying. Since pointing is a prerequisite to establish each link on a certain path, we accounted for the location uncertainty and the availability of a PAT mechanism. Numerical results show that a PAT mechanism along with adaptive optics can provide desirable E2E performance. Thereafter, we developed centralized and distributed routing schemes. For the centralized routing, we proposed variations of shortest path algorithms to optimize E2E rate, BSR, and total power consumption. Finally, we developed a distributed routing protocol LiPaR which combines traveled distance progress and link reliability and manipulates the range-beamwidth tradeoff to find its path. LiPaR is shown to provide better performance than the centralized scheme without a PAT mechanism.
\appendices
\section{Derivations of Communication Ranges}
\label{app:range}
 Range expressions of LoS can be derived by substituting (\ref{eq:Pr_LoS}) into (\ref{eq:rate}). After some algebraic manipulations,  (\ref{eq:Pr_LoS})$\rightarrow$(\ref{eq:rate}) can be put in the form of $c=\frac{1}{x^2}\exp \left \{- \frac{e(\lambda)}{\cos(\phi)} x \right \}$ which has the following root $x=\frac{2 \cos(\phi)}{e(\lambda)}W_0\left( \frac{e(\lambda)}{2\sqrt{c}\cos(\phi)}  \right)$. Notice that this root ($x$) is a solution for the perpendicular distance, hence, the communication range can be obtained from the Euclidean distance $h=x/cos(\varphi)$ where $\varphi$ is the angle between the pointing vector and transmitter-receiver trajectory.  
\section{Convexity Analysis}

\label{app:p1}
Let us start with the convexity analysis of the objective function. Since non-negative weighted sum of convex functions is convex, it is sufficient to prove the convexity of each term, $P_{h-1}^t, \: \forall h \in [1,H]$. $P_{h-1}^t$ is a function of $\erfc^{-1}(2\bar{\mathcal{P}}_{h-1}^h)$  and $\erfc^{-1}(2\bar{\mathcal{P}}_{h-1}^h)^2$. By omitting the hop indices, the second derivative test for these terms can be given as   
\begin{align}
\nonumber \frac{\partial^2 \erfc^{-1}(2\bar{\mathcal{P}})}{\partial \bar{\mathcal{P}}^2} &=2 \pi \erfc^{-1}(2\bar{\mathcal{P}})e^{\left\{ 2\erfc^{-1}(2\bar{\mathcal{P}})^2 \right\}}\\
\nonumber \frac{\partial^2 \erfc^{-1}(2\bar{\mathcal{P}})^2}{\partial \bar{\mathcal{P}}^2} &=2\pi \left(2\erfc^{-1}(2\bar{\mathcal{P}})^2+1\right) e^{ \left\{ 2\erfc^{-1}(2\bar{\mathcal{P}})^2 \right \}}
\end{align}
which are always positive due to the assumption of $\bar{\mathcal{P}} \leq 0.5$. Noting that coefficients in \eqref{eq:DF_Pt} are also non-negative, each term of the objective is a convex function. The mild assumption of $\bar{\mathcal{P}} \leq 0.5$ is also necessary to assure the log-concavity of the Poisson-Binomial distribution. Therefore, we take the logarithm of both sides of the first constraint to put the problem in a convex form. We refer interested readers to \cite{Celik2016Multi,Celik2017Hybrid,Celik2016Green} for a more in-depth convexity analysis of Binomial and Poisson-Binomial distribution. 


\bibliographystyle{IEEEtran}
\bibliography{arxiv}

\begin{thebibliography}{10}
\providecommand{\url}[1]{#1}
\csname url@samestyle\endcsname
\providecommand{\newblock}{\relax}
\providecommand{\bibinfo}[2]{#2}
\providecommand{\BIBentrySTDinterwordspacing}{\spaceskip=0pt\relax}
\providecommand{\BIBentryALTinterwordstretchfactor}{4}
\providecommand{\BIBentryALTinterwordspacing}{\spaceskip=\fontdimen2\font plus
\BIBentryALTinterwordstretchfactor\fontdimen3\font minus
  \fontdimen4\font\relax}
\providecommand{\BIBforeignlanguage}[2]{{%
\expandafter\ifx\csname l@#1\endcsname\relax
\typeout{** WARNING: IEEEtran.bst: No hyphenation pattern has been}%
\typeout{** loaded for the language `#1'. Using the pattern for}%
\typeout{** the default language instead.}%
\else
\language=\csname l@#1\endcsname
\fi
#2}}
\providecommand{\BIBdecl}{\relax}
\BIBdecl

\bibitem{Celik2018Modeling}
A.~Celik, N.~Saeed, T.~Y. Al-Naffouri, and M.-S. Alouini, ``Modeling and
  performance analysis of multihop underwater optical wireless sensor
  networks,'' in \emph{2018 IEEE Wireless Communications and Networking
  Conference (WCNC)}, Apr. 2018, pp. 1--6.

\bibitem{saeed2018survey}
N.~Saeed, A.~Celik, T.~Y. Al-Naffouri, and M.-S. Alouini, ``Underwater optical
  wireless communications, networking, and localization: A survey,''
  \emph{arXiv preprint arXiv:1803.02442}, 2018.

\bibitem{AKYILDIZ2005257}
I.~F. Akyildiz, D.~Pompili, and T.~Melodia, ``Underwater acoustic sensor
  networks: research challenges,'' \emph{Ad Hoc Networks}, vol.~3, no.~3, pp.
  257 -- 279, 2005.

\bibitem{Oubei:15}
H.~M. Oubei, C.~Li, K.-H. Park, T.~K. Ng, M.-S. Alouini, and B.~S. Ooi, ``2.3
  {G}bit/s underwater wireless optical communications using directly modulated
  520 nm laser diode,'' \emph{Opt. Express}, vol.~23, no.~16, pp.
  20\,743--20\,748, Aug 2015.

\bibitem{Shen:16}
C.~Shen, Y.~Guo, H.~M. Oubei, T.~K. Ng, G.~Liu, K.-H. Park, K.-T. Ho, M.-S.
  Alouini, and B.~S. Ooi, ``20-meter underwater wireless optical communication
  link with 1.5 {G}bps data rate,'' \emph{Opt. Express}, vol.~24, no.~22, pp.
  25\,502--25\,509, Oct 2016.

\bibitem{Ballal2015low}
T.~Ballal, T.~Y. Al-Naffouri, and S.~F. Ahmed, ``Low-complexity bayesian
  estimation of cluster-sparse channels,'' \emph{IEEE Trans. Commun.}, vol.~63,
  no.~11, pp. 4159--4173, Nov. 2015.

\bibitem{Kaushal2016underwater}
H.~Kaushal and G.~Kaddoum, ``Underwater optical wireless communication,''
  \emph{IEEE Access}, vol.~4, pp. 1518--1547, 2016.

\bibitem{Saeed2018Connectivity}
N.~Saeed, A.~Celik, T.~Y. Al-Naffouri, and M.-S. Alouini, ``Connectivity
  analysis of underwater optical wireless sensor networks: A graph theoretic
  approach,'' in \emph{IEEE Intl. Conf. Commun. Workshops (ICC Workshops)}, May
  2018, pp. 1--6.

\bibitem{Saeed2018Underwater}
N.~Saeed, A.~Celik, T.~Y. Al-Naffouri, and M.~Alouini, ``Underwater optical
  sensor networks localization with limited connectivity,'' in \emph{IEEE Intl.
  Conf. Acoust, Speech and Signal Process. (ICASSP)}, Apr. 2018, pp.
  3804--3808.

\bibitem{Arnon10underwater}
S.~Arnon, ``Underwater optical wireless communication network,'' \emph{Optical
  Engineering}, vol.~49, pp. 49 -- 49 -- 6, 2010.

\bibitem{Anous2018Vertical}
N.~Anous, M.~Abdallah, M.~Uysal, and K.~Qaraqe, ``Performance evaluation of los
  and nlos vertical inhomogeneous links in underwater visible light
  communications,'' \emph{IEEE Access}, vol.~6, pp. 22\,408--22\,420, 2018.

\bibitem{Hessien2018Experimental}
S.~Hessien, S.~C. Tokgöz, N.~Anous, A.~Boyacı, M.~Abdallah, and K.~A. Qaraqe,
  ``Experimental evaluation of ofdm-based underwater visible light
  communication system,'' \emph{IEEE Photonics Journal}, vol.~10, no.~5, pp.
  1--13, Oct 2018.

\bibitem{Shafique2017Performance}
T.~Shafique, O.~Amin, M.~Abdallah, I.~S. Ansari, M.-S. Alouini, and K.~Qaraqe,
  ``Performance analysis of single-photon avalanche diode underwater vlc system
  using arq,'' \emph{IEEE Photonics Journal}, vol.~9, no.~5, pp. 1--11, Oct
  2017.

\bibitem{Vavoulas2014kconnectivity}
A.~Vavoulas, H.~G. Sandalidis, and D.~Varoutas, ``Underwater optical wireless
  networks: A $k$-connectivity analysis,'' \emph{IEEE J. Ocean. Eng.}, vol.~39,
  no.~4, pp. 801--809, Oct 2014.

\bibitem{Saeed2018performance}
N.~Saeed, A.~Celik, S.~Alouini, and T.~Y. Al-Naffouri, ``Performance analysis
  of connectivity and localization in multi-hop underwater optical wireless
  sensor networks,'' \emph{To appear in IEEE Transactions on Mobile Computing},
  pp. 1--1, 2018.

\bibitem{Akhoundi2015cellular}
F.~Akhoundi, J.~A. Salehi, and A.~Tashakori, ``Cellular underwater wireless
  optical {CDMA} network: Performance analysis and implementation concepts,''
  \emph{IEEE Trans. Commun.}, vol.~63, no.~3, pp. 882--891, March 2015.

\bibitem{Akhoundi2016cellular}
F.~Akhoundi, M.~V. Jamali, N.~B. Hassan, H.~Beyranvand, A.~Minoofar, and J.~A.
  Salehi, ``Cellular underwater wireless optical {CDMA} network: Potentials and
  challenges,'' \emph{IEEE Access}, vol.~4, pp. 4254--4268, 2016.

\bibitem{Jamali2016performance}
M.~V. Jamali, F.~Akhoundi, and J.~A. Salehi, ``Performance characterization of
  relay-assisted wireless optical {CDMA} networks in turbulent underwater
  channel,'' \emph{IEEE Trans. Wireless Commun.}, vol.~15, no.~6, pp.
  4104--4116, June 2016.

\bibitem{Jamali2017multihop}
M.~V. Jamali, A.~Chizari, and J.~A. Salehi, ``Performance analysis of multi-hop
  underwater wireless optical communication systems,'' \emph{IEEE Photon.
  Technol. Lett.}, vol.~29, no.~5, pp. 462--465, March 2017.

\bibitem{Elamassie2018performance}
M.~Elamassie, F.~Miramirkhani, and M.~Uysal, ``Performance characterization of
  underwater visible light communication,'' \emph{IEEE Trans. Commun.}, pp.
  1--1, 2018.

\bibitem{Kahn97wireless}
J.~M. Kahn and J.~R. Barry, ``Wireless infrared communications,'' \emph{Proc.
  IEEE}, vol.~85, no.~2, pp. 265--298, Feb 1997.

\bibitem{Saeed2018Robust}
N.~Saeed, A.~Celik, T.~Y. Al-Naffouri, and M.-S. Alouini, ``Robust {3D}
  localization of underwater optical wireless sensor networks via low rank
  matrix completion,'' in \emph{IEEE 19th Intl. Workshop Signal Process. Adv.
  Wireless Commun. (SPAWC)}, Jun. 2018, pp. 1--5.

\bibitem{Saeed2017Energy}
N.~Saeed, A.~Celik, T.~Y. Al-Naffouri, and M.~Alouini, ``Energy harvesting
  hybrid acoustic-optical underwater wireless sensor networks localization,''
  \emph{Sensors}, vol.~18, no.~1, 2017.

\bibitem{Arnon09nonLOS}
S.~Arnon and D.~Kedar, ``Non-line-of-sight underwater optical wireless
  communication network,'' \emph{J. Opt. Soc. Am. A}, vol.~26, no.~3, pp.
  530--539, Mar 2009.

\bibitem{fernandez2010closed}
M.~Fernandez and S.~Williams, ``Closed-form expression for the poisson-binomial
  probability density function,'' \emph{IEEE Trans. Aerosp. Electron. Syst.},
  vol.~46, no.~2, pp. 803--817, April 2010.

\bibitem{agrawal2012fiber}
G.~P. Agrawal, \emph{Fiber-{O}ptic {C}ommunication {S}ystems}.\hskip 1em plus
  0.5em minus 0.4em\relax John Wiley \& Sons, 2012, vol. 222.

\bibitem{karp1988optical}
S.~Karp, R.~Gagliardi, S.~Moran, and L.~Stotts, ``Optical {C}hannels:
  {F}ibers,'' \emph{Clouds, Water, and the Atmosphere: Plenum Press New York},
  1988.

\bibitem{kaibel2006bottleneck}
V.~Kaibel and M.~A. Peinhardt, \emph{On the {B}ottleneck {S}hortest {P}ath
  {P}roblem}.\hskip 1em plus 0.5em minus 0.4em\relax Konrad-Zuse-Zentrum
  f{\"u}r Informationstechnik, 2006.

\bibitem{KSP}
D.~Eppstein, ``Finding the k shortest paths,'' \emph{SIAM Journal on
  Computing}, vol.~28, no.~2, pp. 652--673, 1998.

\bibitem{Celik2016Multi}
A.~Celik and A.~E. Kamal, ``Multi-objective clustering optimization for
  multi-channel cooperative spectrum sensing in heterogeneous green crns,''
  \emph{IEEE Transactions on Cognitive Communications and Networking}, vol.~2,
  no.~2, pp. 150--161, Jun. 2016.

\bibitem{Celik2017Hybrid}
A.~Celik, A.~Alsharoa, and A.~E. Kamal, ``Hybrid energy harvesting-based
  cooperative spectrum sensing and access in heterogeneous cognitive radio
  networks,'' \emph{IEEE Transactions on Cognitive Communications and
  Networking}, vol.~3, no.~1, pp. 37--48, Mar. 2017.

\bibitem{Celik2016Green}
A.~Celik and A.~E. Kamal, ``Green cooperative spectrum sensing and scheduling
  in heterogeneous cognitive radio networks,'' \emph{IEEE Transactions on
  Cognitive Communications and Networking}, vol.~2, no.~3, pp. 238--248, Sept
  2016.

\end{thebibliography}

\end{document}